\documentclass[aps,prd,groupedaddress,nofootinbib,preprintnumbers]{revtex4}

\usepackage{graphicx}
\usepackage{latexsym}
\usepackage{amsfonts}
\usepackage{amssymb}
\usepackage{amsmath}
\usepackage{slashed}
\usepackage{tabularx,url,color}
\usepackage{hyperref}
\usepackage{multirow}

\newcommand{\gev}{{\ensuremath \text{GeV}}}

\newcommand{\ifb}{{\ensuremath \text{fb}^{-1}}}
\newcommand{\xpass}{${\mathfrak t_w}$}
\newcommand{\xfail}{${\mathfrak t_{-}}$}

\def\eg{{\sl e.g.} \,}
\def\ie{{\sl i.e.}}
\def\etal{{\sl et al} \,}

\begin{document}

\title{Buckets of Tops}

\author{Matthew R.~Buckley$^1$, Tilman Plehn$^2$, and Michihisa Takeuchi$^3$}

\affiliation{$^1$Center for Particle Astrophysics, Fermi National Accelerator Laboratory, Batavia, IL, U.S.}
\affiliation{$^2$Institut f\"ur Theoretische Physik, Universit\"at Heidelberg, Germany}
\affiliation{$^3$Theoretical Physics and Cosmology Group, Department of Physics, King's College London, London~WC2R 2LS, UK}
\date{\today}

\begin{abstract}
  Reconstructing hadronically decaying top quarks is a key challenge
  at the LHC, affecting a long list of Higgs analyses and new physics
  searches. We propose a new method of collecting jets in buckets,
  corresponding to top quarks and initial state radiation. This method
  is particularly well suited for moderate transverse momenta of the
  top quark, closing the gap between top taggers and traditional top
  reconstruction. Applying it to searches for supersymmetric top
  squarks we illustrate the power of buckets.
\end{abstract}

\maketitle

\section{Introduction}
\label{sec:intro}

An important difference between the Tevatron and LHC is that the latter
can produce and study top quarks in great
numbers~\cite{tops_ex}. This allows us to investigate all different top
production mechanisms in detail, including their QCD
structure. After the discovery of a Higgs-like~\cite{higgs} resonance, studying
its coupling to the top quark will play a particularly critical role in our
understanding of the Higgs sector. This is made most obvious in the
renormalization group evolution of the Higgs potential to large energy
scales~\cite{lecture}. The direct measurement of the top Yukawa
coupling clearly hinges on top quark identification and
reconstruction.  At the same time, we have reason to suspect that
new physics that solves the hierarchy problem and lives
at sufficiently high energy scales tends to couple
strongly to top quarks~\cite{bsm_review}. This motivates us to search for
new physics in the LHC top sample; for example by searching for top pair
resonance structures of top pair production associated with missing
transverse momentum.

Historically, the study of top pair production has largely been
restricted to semi-leptonic decays of the two tops quarks. The reason
is that the lepton effectively removes the overwhelming QCD
background. However, purely leptonic top pairs not only come at a much smaller
rate, they also include two neutrinos, challenging any analysis based
on the observed missing transverse momentum. A major challenge in top
physics at the LHC is how to gain access to the purely hadronic decays
of top quarks.\bigskip

Identifying hadronic top decays using a jet algorithm was part of the
original proposal of jet substructure
analyses~\cite{seymour,tagger_review}. Some of the early jet
substructure algorithms were designed to target hadronic top
decays~\cite{early_toptagger}. While Higgs taggers~\cite{bdrs} should
clearly have a high priority within the LHC experiments, working top
taggers are the perfect laboratory to test how well substructure
approaches work in practice. 

The moment we go beyond searches for heavy resonances the main problem
of all top taggers is the size of the initial fat jet. For example, using
the Cambridge--Aachen jet of size $R=1.5$ as the starting point of the
{\sc HEPTopTagger}~\cite{tth,HEP1,atlas} limits the momentum range of
reconstructable top quarks to $p_{T,t} \gtrsim 200$~GeV. Essentially,
all other top tagging approaches require even higher boost.
Increasing the size of the fat jet to $R=1.8$ raises several QCD and
combinatorics issues~\cite{HEP3}. The big question in using hadronic top
analyses as part of Higgs searches or top partner searches 
is how to further reduce this top momentum threshold.\bigskip


In this paper we propose an alternative method for an efficient top
reconstruction at moderate momentum.  It targets the transverse
momentum regime, 
\begin{equation} 
 p_{T,t} = 100 - 350~\gev \; ,
\end{equation}
in the fully hadronic decay mode. Starting from an event with a high
multiplicity of jets, we assign all jets into three groups or
`buckets'. The buckets are chosen based on a metric in terms of
invariant masses, defining two top buckets and a third bucket
containing the extra hadronic activity like initial state radiation (ISR). While
initially this search strategy does not prefer boosted top quarks, we
will see how such events are eventually preferred from a combinatorics
perspective.\bigskip

In Section~\ref{sec:simple} we start with a simple algorithm for
reconstructing tops in buckets. We test this algorithm for
hadronically decaying top pairs as well as $W$+jets and pure QCD jets
backgrounds. Additional handles will help us separate the top signal
from the backgrounds.  In Section~\ref{sec:bb}, we modify the simple
algorithm to take advantage of the $b$ quarks and $W$ bosons that are
present in top decays but not in the QCD backgrounds. This improved
bucket algorithm is optimized to efficiently find and reconstruct top
pairs with moderate $p_T$.  In Section~\ref{sec:stop} we apply our
bucket algorithm to stop pair searches.

\section{Simple Bucket Algorithm}
\label{sec:simple}

In this section, we start with a simple algorithm to identify and
reconstruct hadronically decaying top pairs.  While an improved algorithm will be presented in the next section, this simple version captures many of the key concepts we will employ later.
The overall scheme is
fairly straightforward:
by assumption every jet originates from one of the two tops or from
initial state radiation, so we assign every jet to one of three
`buckets'. Jets in buckets $B_1$ and $B_2$ correspond to top decays,
while all remaining jets are placed in $B_\text{ISR}$. We cycle
through every permutation of jet assignments to minimize the distance
between the invariant masses of the jets in $B_1$ and $B_2$ and the
top mass. The metric is chosen to ensure that bucket $B_1$
reconstructs the top mass better than bucket $B_2$.\bigskip

Here and throughout the remainder of the paper, all Standard Model (SM) samples are generated with {\sc Alpgen+Pythia}~\cite{alpgen,pythia}. 
We use matrix-level matching~\cite{mlm} to correctly describe jet
radiation over the entire phase space. This includes up to
$t\bar{t}$+2~jets, $W$+4~jets and $3-5$ QCD jets, with the top cross
sections normalized to next-to-next-to-leading order~\cite{top_nnlo}.
Jets are reconstructed using the Cambridge/Aachen
algorithm~\cite{ca_algo} of size $R=0.5$ in {\sc
  FastJet}~\cite{fastjet}. Note that all our results are relatively
insensitive to the choice of jet algorithm.

All leptons we require to be hard and isolated: $p_{T,\ell} > 10$ GeV
and no track of another charged particle within $R < 0.5$ around the
lepton.  We consider only jets with $p_T>25$~GeV and
$|\eta|<2.5$. 
Even though the algorithm presented in this section is in principle applicable to
events with any number of jets we preselect events with five or more
jets to reduce QCD backgrounds. Because we are interested in
hadronically decaying $t\bar{t}$ pairs we veto on isolated
leptons. The restricted sample denoted as $t_h\bar{t}_h$ has a cross
section of 104~pb at the LHC with $\sqrt{s} = 8$~TeV. One last word
concerning underlying event and pile-up: unlike methods involving jet 
substructure~\cite{tagger_review} our bucket reconstruction relies on
standard jets with moderately large multiplicities, so aside from jet
energy scale uncertainties we do not expect specific experimental or
theoretical challenges.

\subsection*{Bucket definition}

As the goal of the bucket algorithm is to identify tops by sorting
jets into categories that resemble tops, we need a metric to determine
the similarity of a collection of jets to a top. For simple buckets
$B_i$ it is
\begin{equation}
\Delta_{B_i} = |m_{B_i} - m_t| 
\qquad \text{with} \qquad 
m_{B_i}^2= \left( \sum_{j \in B_i} p_j \right)^2 \; , 
\label{eq:delta}
\end{equation}
where we sum over all four-vectors in the bucket. For each event with
five or more jets we permute over all possible groupings of the jets
into three buckets $\{B_1,B_2,B_\text{ISR}\}$. We then select the
combination that minimizes a global metric defined as
\begin{equation}
\Delta^2=\omega \Delta_{B_1}^2 + \Delta_{B_2}^2 \; .
\label{eq:deltaevent}
\end{equation}
The factor $\omega>1$ stabilizes the grouping of jets into buckets. In this work we take
$\omega= 100$, effectively decoupling $\Delta_{B_2}$ from the metric. As a
consequence we always find $\Delta_{B_1} < \Delta_{B_2}$, \ie~$B_1$ is
the bucket with an invariant mass closer to that of the top than the
invariant mass of bucket $B_2$. Other values of $\omega$ might eventually
turn out more appropriate for different applications.\bigskip

As the first selection cut we require the invariant masses of both top
buckets, $B_1$ and $B_2$, to lie in the window
\begin{equation}
155~\gev < m_{B_{1,2}} < 200~\gev \; .
\label{eq:mBcut}
\end{equation}
The lower limit selects events above the Jacobian peak for top
decays. We will see that this selection improves the top signal over
QCD background $S/B$ by about a factor of two.  All buckets passing
Eq.~\eqref{eq:mBcut} we categorize by their number of jets; buckets
including three or more jets ($3j$-buckets) and those including two
jets ($2j$-buckets). Selecting only events with two $3j$-buckets
improves the signal-to-background ratio by a factor of five.

\subsection*{Jet selection}
\label{sec:2}

For tagging two tops in fully hadronic mode, we might naively require
at least six reconstructed jets. In practice, with a threshold of
$p_{T,j} > 25$~GeV this condition is too strict. To improve our
efficiency we need to consider the case where one of the jets from top
pair decays is missing. It is also worth noting that even requiring six
jets does not guarantee that we collect all six decay products of the
top pair. Frequently, some of the observed jets come from
initial state radiation instead~\cite{HEP3}.\bigskip

\begin{figure}[t]
\includegraphics[width=0.32\textwidth]{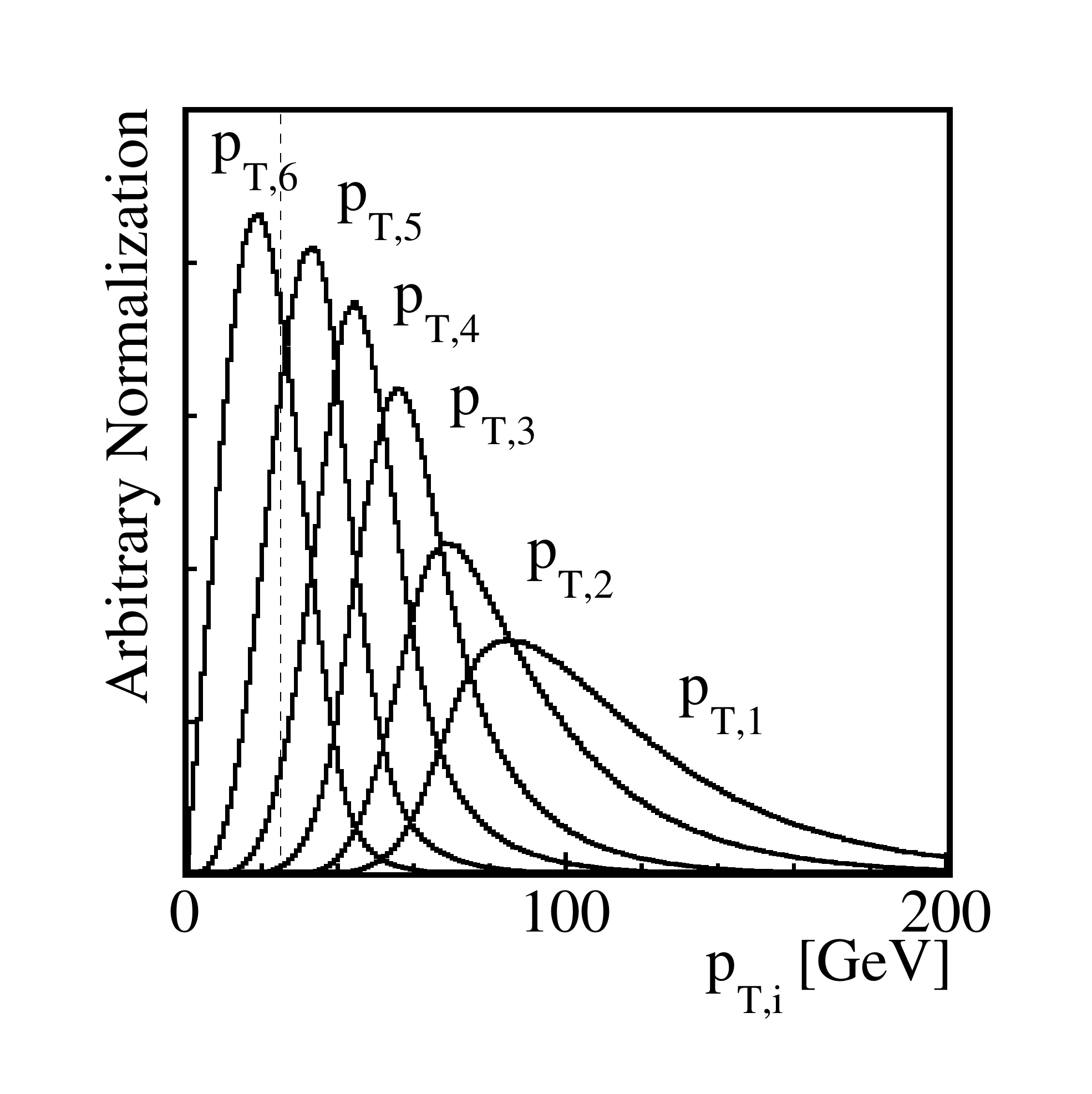}
\includegraphics[width=0.32\textwidth]{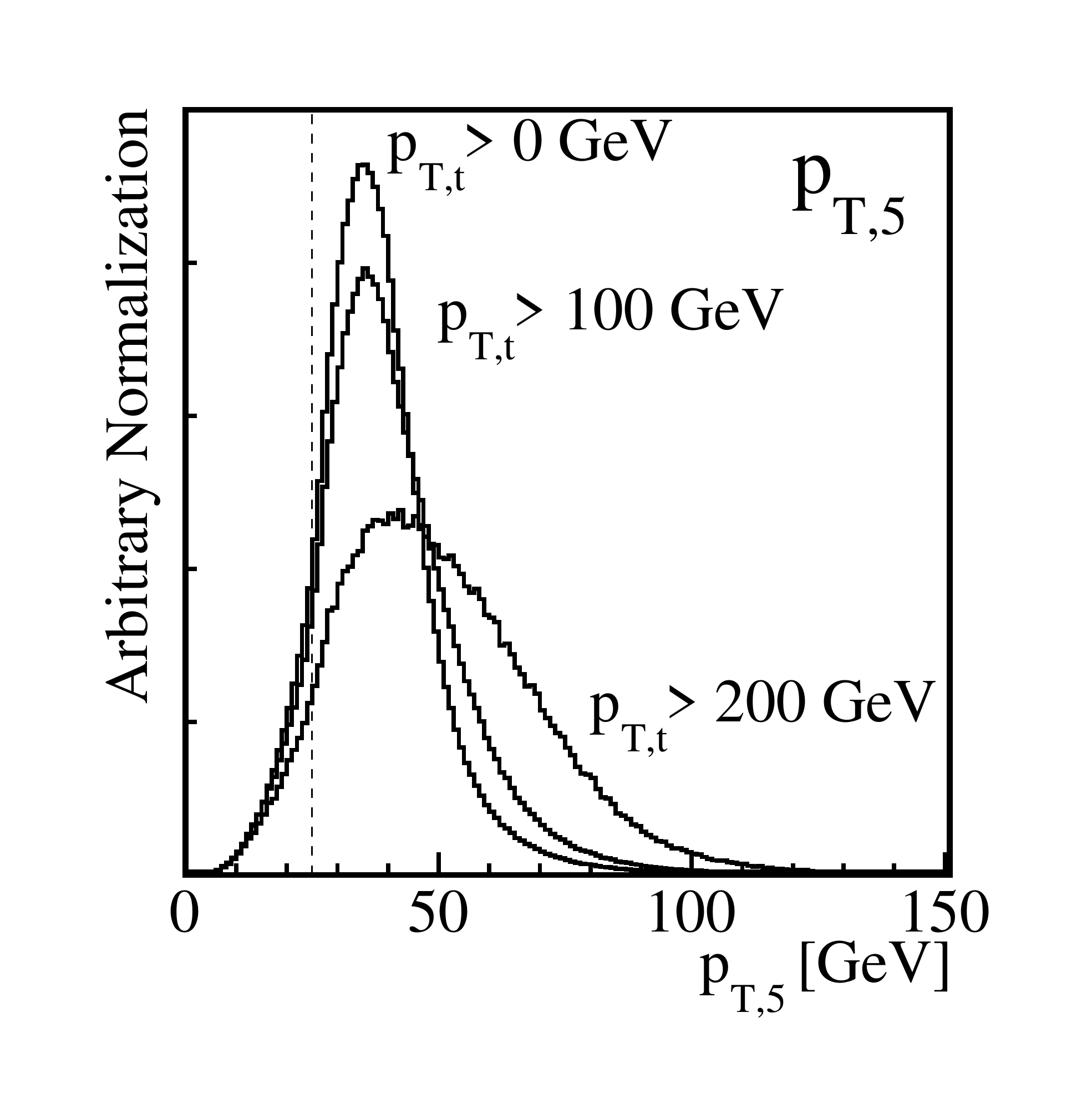}
\includegraphics[width=0.32\textwidth]{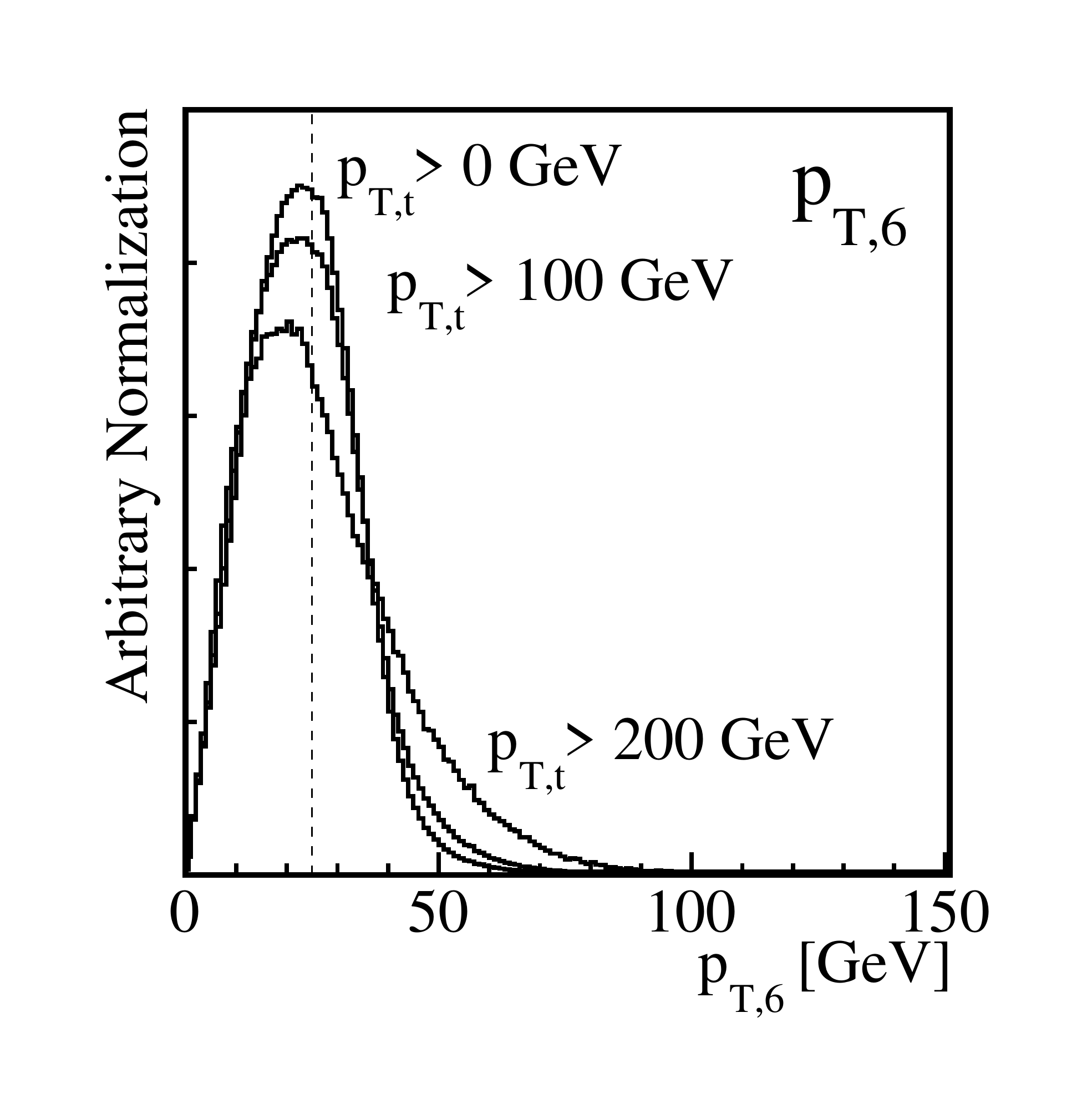}
\caption{Normalized transverse momentum distributions for the top
  decay partons in the $t_h \bar{t}_h$ sample.  Left: all six
  $p_{T,i}$ distributions. Central and right: normalized distributions of $5^\text{th}$
  and $6^\text{th}$ hardest partons for events with at least 5 jets. 
  Different lines in the central and
  right panels correspond to the different generator-level cuts on the
  top transverse momenta $p_{T,t} > 0, 100, 200$~GeV.}
\label{fig:pt}
\end{figure}

In Figure~\ref{fig:pt} we plot the parton level $p_T$ distributions of
the six decay partons from the top pairs. In the left panel we see
that the four hardest decay jets are not affected by the threshold
$p_{T,j}> 25$~GeV.  In contrast, the softest distribution only peaks
around 25~GeV, so roughly half the events do not pass our threshold on
the sixth jet. 

\begin{table}[b]
\centering
\begin{tabular}{ll||c|c|c}
\hline
&& $t_h\bar{t}_h$+jets [pb] &\ \  $p_{T,6}>25$~GeV &\ \ \ \  $p_{T,5}>25~\gev> p_{T,6}$ \\
\hline
lepton veto &&  104.1 & 33.4\% & 44.9\% \\
\hline
$n_j\ge 4$ && 94.3 & 35.8\% & 46.0\% \\
$n_j\ge 5$ &&70.5& 42.5\% &   46.4\% \\ 
$n_j\ge 6$ &&  36.7&  54.7\% &   38.0\%\\
\hline
\multirow{2}{*}{$n_j\ge 5$} & $p_{T,t_2}>100~\gev$ & 32.7 & 43.6\% & 46.2\% \cr
                            & $p_{T,t_2}>200~\gev$ & 6.7 & 47.4\% & 44.7\% \cr
\hline
\end{tabular}
\caption{Signal cross sections after requiring five or six top decay
  jets with $p_{T,j} > 25$~GeV. The reference value are all hadronic
  top pairs after applying the lepton veto as described in the text.}
\label{tab:partons}
\end{table}

Table~\ref{tab:partons} shows the number of events in the hadronic
$t_h\bar{t}_h$ sample after several cuts on the jet multiplicity, and
the percentage of events with the 5$^\text{th}$ or $6^\text{th}$
parton-level top decay jets above $p_{T,j} > 25~\gev$. 
In about a half of events with at least six jets the sixth top-decay parton
falls below the $p_T$ threshold. Adding the two columns tells us
that more than 90\% of all events capture five of the six
top decay products. Requiring only five instead of six jets
increases the fraction of events where we miss only one of the top decay
products to almost half. The table also shows the effect of placing a
transverse momentum cut on the softer top, $p_{T,t_2}$.  For a
moderate top $p_T$ threshold our central values for the efficiencies
are not strongly affected, but hadronization as well as detector
effects might lead to significant shifts due to the steep $p_{T,j}$
behavior.

\subsection*{W reconstruction}

After placing each of the jets in the event into one of three buckets
($B_1$, $B_2$, or $B_\text{ISR}$) we require the $3j$-buckets to
contain a hadronically decaying $W$ candidate.  In the rare case of
one bucket consisting of more than three jets we combine them into
exactly three jets using the C/A algorithm and then look for a $W$
candidate.  As in the {\sc HEPTopTagger}~\cite{HEP1} we define a mass
ratio cut
\begin{equation}
\left|\frac{m_{k\ell}}{m_{B_i}} - \frac{m_W}{m_t}\right| <0.15
\label{eq:Wcut}
\end{equation}
for at least one combination of jets $k,\ell$ in the bucket $i$.
Events with $2j$-buckets by construction cannot satisfy Eq.~\eqref{eq:Wcut}. 
In addition, in such events one of the $W$ decay jets is typically 
the softest jet and does not pass the $p_T$ threshold, and so the $W$
reconstruction could not occur regardless\bigskip

In our first, naive approach we
categorize all events with two valid top buckets into three
types:
\begin{itemize}
\item (\xpass,\xpass): both top buckets have $W$ candidates as defined
  by Eq.~\eqref{eq:Wcut},
\item (\xpass,\xfail) or (\xfail,\xpass): only the first or second top
  bucket has a $W$ candidate,
\item (\xfail,\xfail): neither top bucket has a $W$ candidate.
\end{itemize}
The \xpass\, or \xfail\, status is ordered as $(B_1,B_2)$, where $B_1$ is
defined as the bucket closest in mass to the top. Buckets classified as
\xpass\,  have to be $3j$-buckets, while \xfail\, buckets can be either
$3j$ or $2j$.

To extract hadronic top pair events from the QCD background we can
compare the different categories on Monte-Carlo truth level. Starting
from $S/B \sim 0.005$ after the lepton veto and selecting only
(\xpass,\xpass) events yields the highest value $S/B \sim 0.09$. This
corresponds to an improvement of $S/B$ by almost a factor 20. To
improve beyond this level, we need to require at least one, preferably
two, $b$-tags to control the mostly Yang--Mills and light-flavor QCD
background.

\subsection*{b-tags}

To further reduce the QCD background we exploit $b$-tags. We assume
$b$-tagging and mis-tagging efficiencies for light flavors
($\epsilon_b, \epsilon_\text{mis}$) to be (70\%, 1\%), and fully
account for combinatorial factors in the background. 
For the $t\bar{t}$+jets signal the effect of mis-tagging is
sub-leading and can be ignored.

To avoid combinatorics we could impose $b$-tagging only for the most
likely $b$-jet in a bucket based on the $W$ condition factors to
improve $S/B$, as suggested in Ref.~\cite{HEP3}. In this algorithm we
do not take this option because it reduces the signal efficiency. We
prefer to keep the maximum fraction of signal events especially for
the case that both signal and the main background include $t\bar{t}$
events, such as the top partner searches discussed below.\bigskip

In any top-tagging algorithm we are interested not only in extracting
the signal from backgrounds, but in accurately reconstructing the
original top momenta. For a measure of our reconstruction accuracy we use the geometric distance in
the ($\eta, \phi$) plane between the bucket momentum and
the closer top parton momentum $p_t$ obtained from Monte-Carlo truth,
\begin{equation}
R_i = \min[\Delta R(B_i, p_{t_1}),\Delta R(B_i, p_{t_2})] \; .
\end{equation}
We consider our reconstruction successful when $R_i < 0.5$. In the
following, we indicate the percentage of events with both buckets
reconstructing top momenta ($R_1 < 0.5$ and $R_2 < 0.5$), events where only $B_1$
reconstructs the top momentum ($R_1 < 0.5 < R_2$), and events where only $B_2$
reconstructs the top momentum ($R_2 < 0.5 < R_1$). The last case
allows for events where the second bucket (with its worse top mass
reconstruction) actually gives the better top direction.\bigskip

For (\xpass,\xpass) events where each bucket contains exactly one $b$-jet,
the top momentum reconstruction is generally good.
As seen in to Table~\ref{tab:nj_bimposed} about 75\% of the events
reconstruct both top directions well.  
As expected from the discussion above, a significant
fraction of signal events only give ($3j$,$2j$)-buckets.  When a
$W$ candidate is not found, but each bucket contains a $b$-tag and
lies in the top mass window Eq.~\eqref{eq:mBcut}, the momentum
reconstruction is good only for the \xpass\, bucket; in these events half of the \xpass\, momenta with a
$W$ candidate reconstruct the top direction well.
All this points to using the $b$-tag information to improve our
reconstruction algorithm.  This will be the starting point of the
improved algorithm in the next section.

\begin{table}[t]
\centering
\begin{tabular}{l||r|r|r|r||r|r|r}
\hline
&  $t_h\bar{t}_h$+jets [fb] & $ R_{1}, R_{2}<0.5$  & $ R_{1}<0.5< R_{2}$ & $ R_{2}<0.5<R_{1} $ &QCD [fb] & $W$+jets [fb] & $S/B_\text{QCD}$ \\
\hline
lepton veto  & 104$\times 10^3$ & -- & -- & -- & -- & -- &  \\
\hline
5 jets or more  &  70.3$\times 10^3$ & &&& 14643$\times 10^3$ & 96.7$\times 10^3$  & 0.005 \\ 
$m_{B}$ cut [Eq.(\ref{eq:mBcut})] & 36.1$\times 10^3$ &  &&& 4768$\times 10^3$ & 34.4$\times 10^3$ & 0.008 \\
\hline
(\xpass,\xpass), 2$b$-tag & 1811 & 74.5\% & 6.9\% & 6.0\% & 63.2 & 0.58 & 28.65\cr
\hline
(\xpass,\xfail), 2$b$-tag  & 1513 & 26.8\% & 24.6\% & 7.3\% & 466 & 3.77 & 3.25 \cr
(\xfail,\xpass), 2$b$-tag  & 1066 & 26.5\% & 8.3\%  & 21.1\% & 362 & 2.78 & 2.95 \cr
(\xfail,\xfail), 2$b$-tag  & 1615 & 9.9\% & 13.3\% & 14.7\%  & 1348 & 8.55 & 1.20 \cr
\hline
\end{tabular}
\caption{Numbers of events for simple buckets with one $b$-tag on each bucket, passing
  various levels of top reconstruction, as described in the
  text. Events that do not reconstruct tops to within $R_1,R_2 < 0.5$
  are not shown, but consist of the remaining percentage of events for
  each category. 
  }
\label{tab:nj_bimposed}
\end{table}

\section{Bottom-centered buckets}
\label{sec:bb}

In Section~\ref{sec:simple} we have seen that we need at least two
$b$-tags per event to control the QCD background. However, in the simple algorithm, each
bucket does not always have exactly one $b$-jet and the reconstruction
is not particularly effective for (\xpass,\xpass) events. The
obvious solution is to define buckets around $b$-tagged jets, \ie~starting 
each bucket with the bottom jets (which are usually the hardest jets in the event) and adding
light-flavor jets to it.\bigskip

In this section we define buckets starting with the requirement that $B_1$ and $B_2$ each have
exactly one $b$-jet, and restrict the possible permutation of jet assignments to $B_1$, $B_2$, and $B_\text{ISR}$ accordingly.
 Other than this, we use the same distance
measure defined in Eq.~\eqref{eq:delta} and Eq.~\eqref{eq:deltaevent}
and select $\{B_1, B_2, B_\text{ISR}\}$ giving the minimum
$\Delta$.

\begin{figure}[b]
\includegraphics[width=0.3\textwidth]{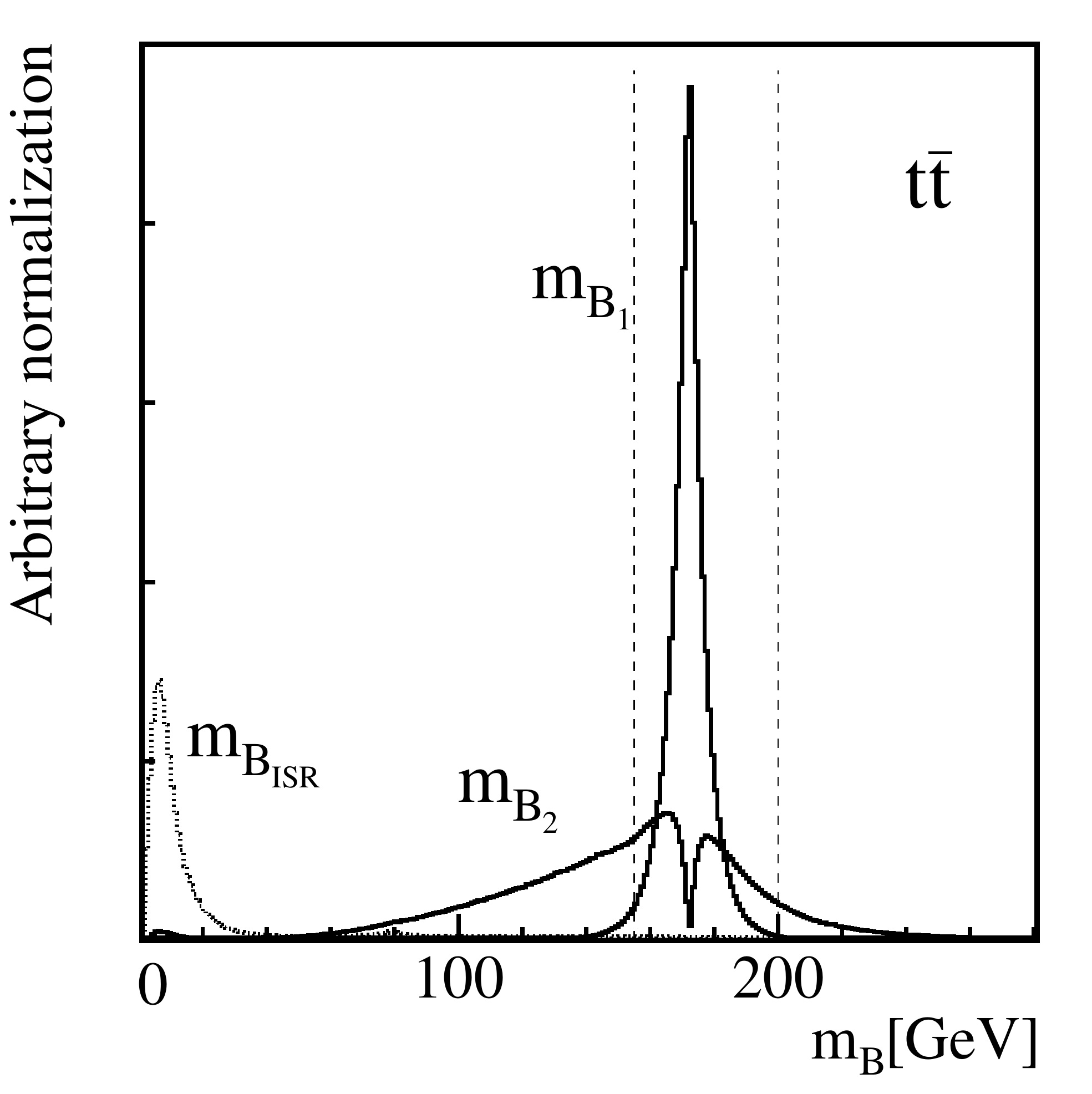}
\hspace*{0.15\textwidth}
\includegraphics[width=0.3\textwidth]{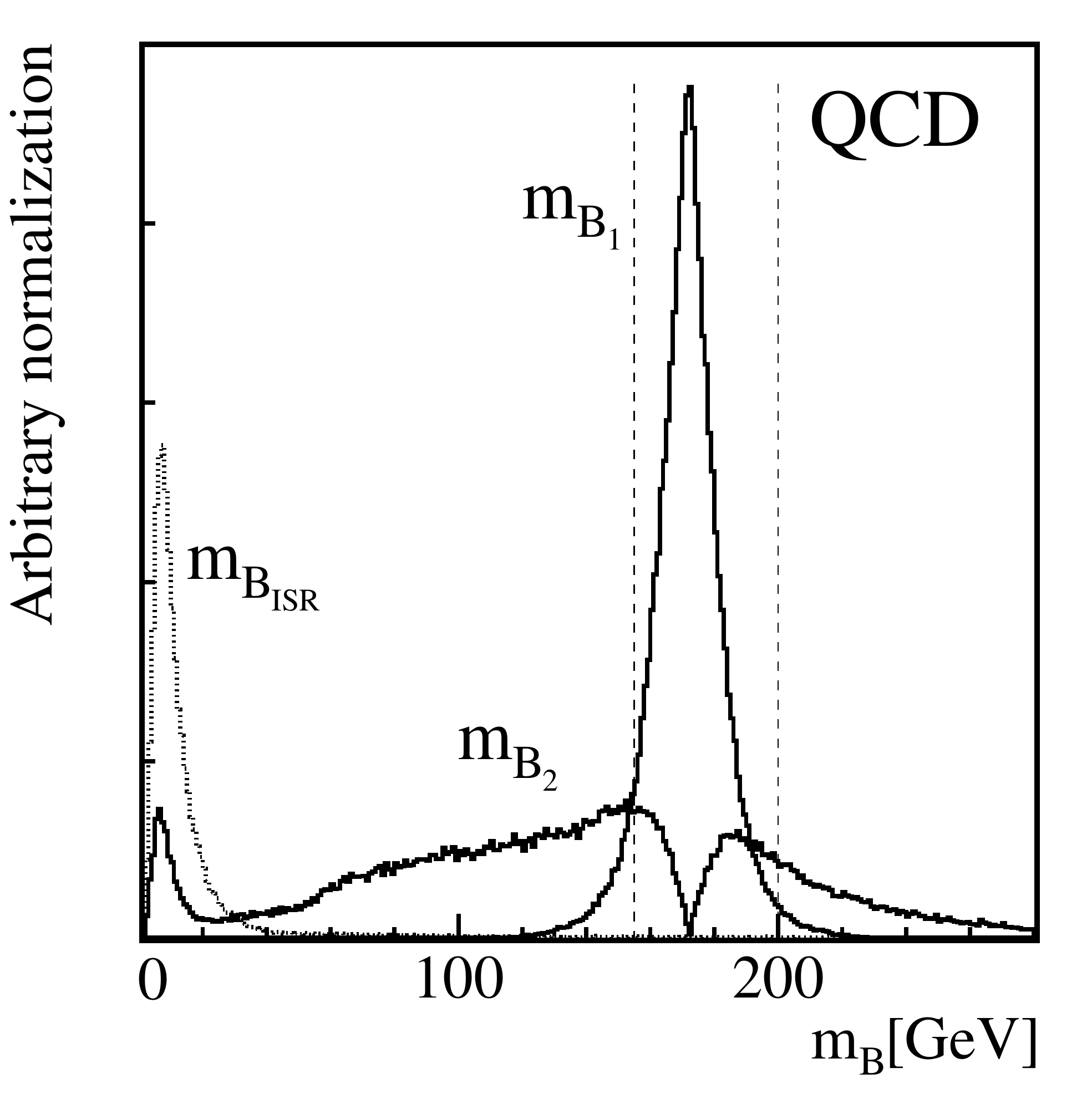}
\caption{Bucket masses for $t\bar{t}$ (left) and QCD (right)
  events. We select all events with $n_j \geq 5$, skipping the mass
  window for the top buckets.}
\label{fig:bucketmass}
\end{figure}

Figure~\ref{fig:bucketmass} shows the bucket masses $m_{B_1}$,
$m_{B_2}$ and $m_\text{ISR}$. For both $t\bar{t}$ and QCD samples the
$m_{B_1}$ distributions peak at $m_t$ by construction. The
distribution is narrower for the signal. The dip in the $m_{B_2}$
distributions at $m_t$ is due to the large weighting factor $\omega$ in
Eq.~\eqref{eq:deltaevent}, which defines the bucket with mass
closest to $m_t$ to be $B_1$. Compared with the $m_{B_1}$
distributions, the $m_{B_2}$ distributions are broad but still tend to
peak toward $m_t$.\bigskip

As mentioned above, the analysis of the top buckets constructed around
the $b$-tagged jets is the same as the simple algorithm described in
Section~\ref{sec:simple}, including the bucket mass cut in in
Eq.~\eqref{eq:mBcut}.  In Table~\ref{tab:nj_bimposed_summary} we show
the corresponding results.  Starting with two $b$-jets improves the
number of (\xpass,\xpass) events by almost 50\%.  Roughly 70\% of (\xpass,
\xpass)-events reconstruct both tops well, essentially unchanged from
the earlier analysis.  One kind of events which is now correctly
accounted for are cases where the simple algorithm finds two $b$-jets
in the same bucket, and give a bucket mass in the correct range.

Asking for two $b$-tags within at least five jets at the very
beginning produces large combinatorial factors for mis-tagging QCD
background events. As a result the backgrounds double in each category
and $S/B$ degrades for (\xpass,\xpass) events.\bigskip

\begin{table}[t]
\centering
\begin{tabular}{l||r|r|r|r||rr|r}
\hline
 &  $t_h\bar{t}_h$+jets [fb] & $R_{1}, R_2<0.5$ &$R_{1}<0.5$ & $R_{2}<0.5$ &QCD [fb] & $W$+jets [fb] & $S/B_\text{QCD}$  \\
\hline
5 jets, 2$b$-tag &21590& &&& 16072 &    109.6& 1.36  \\
\hline
(\xpass,\xpass)  & 2750 & 68.9\% & 9.3\% & 7.5\% & 126.2 & 1.181 & 21.8 \\
\hline
(\xpass,\xfail)  & 2517 & 23.4\% & 25.6\% & 8.5\% & 727.1 & 6.03 & 3.5 \\
(\xfail,\xpass)  & 1782 & 21.8\% & 9.1\% & 22.6\% & 596.5 & 4.85 & 3.0 \\
(\xfail,\xfail) & 2767 & 9.0\% & 14.3\% & 13.9\% & 2002 & 14.05 & 1.4 \\
\hline
\end{tabular}
\caption{Signal and background rates passing various levels of
  reconstruction, requiring one $b$-jet in each top buckets $B_{1,2}$.
  The base-line selection cuts are the same as in
  Table~\ref{tab:nj_bimposed}.}
\label{tab:nj_bimposed_summary}
\end{table}

While there is no obvious way to improve the (\xpass,\xpass) category of
events, Table~\ref{tab:nj_bimposed_summary} shows that a significant
number of events come out (\xpass,\xfail) and (\xfail,\xpass), that is, only one bucket
contains a $W$ candidate. For these events, the QCD background is
not huge, $S/B\sim 3$, so we will try to improve our treatment of this
fraction of events.

\subsection*{b/jet Buckets}
\label{sec:bj}

\begin{figure}[t]
\includegraphics[width=0.32\textwidth]{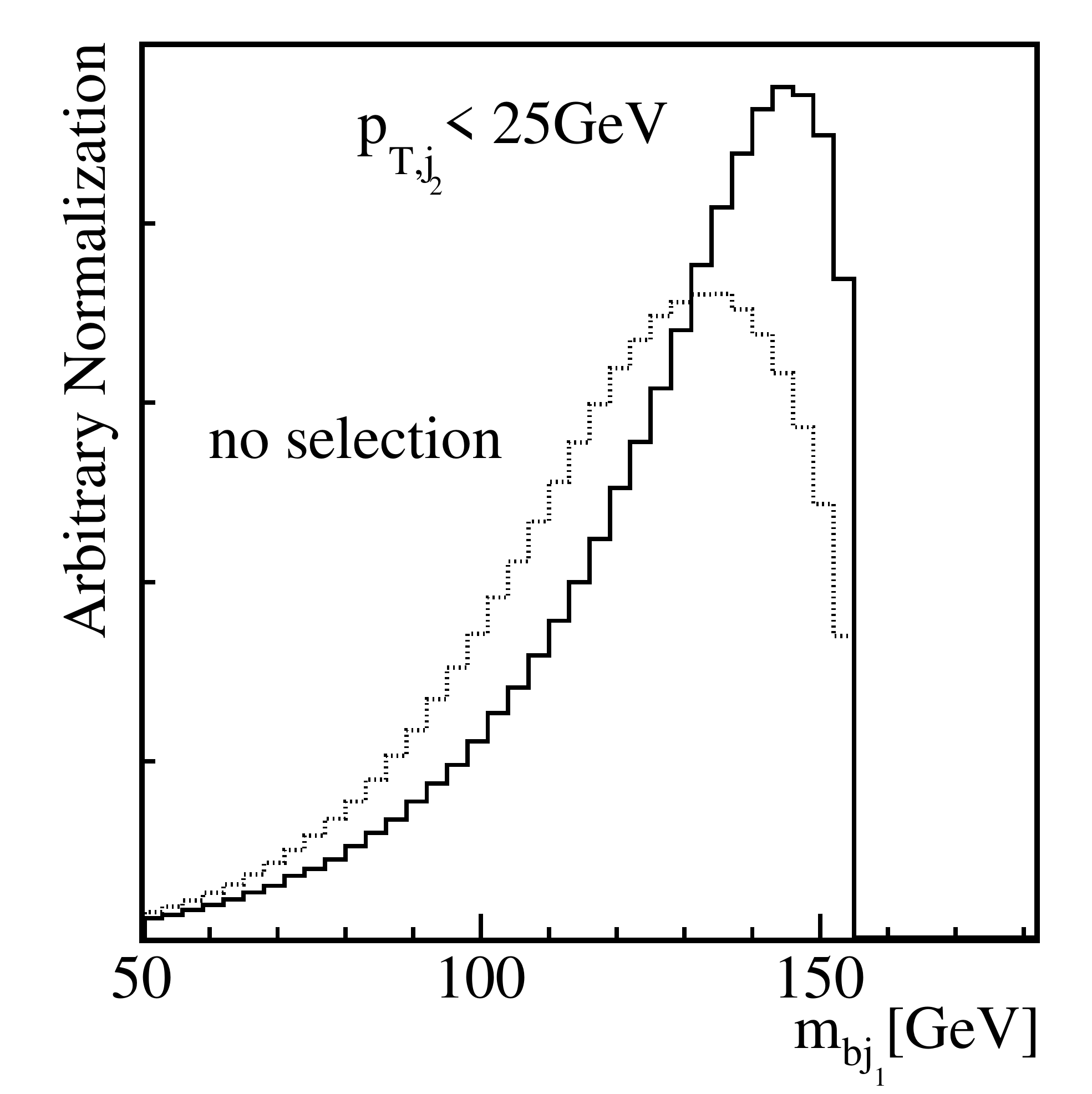}
\hspace*{0.15\textwidth}
\includegraphics[width=0.32\textwidth]{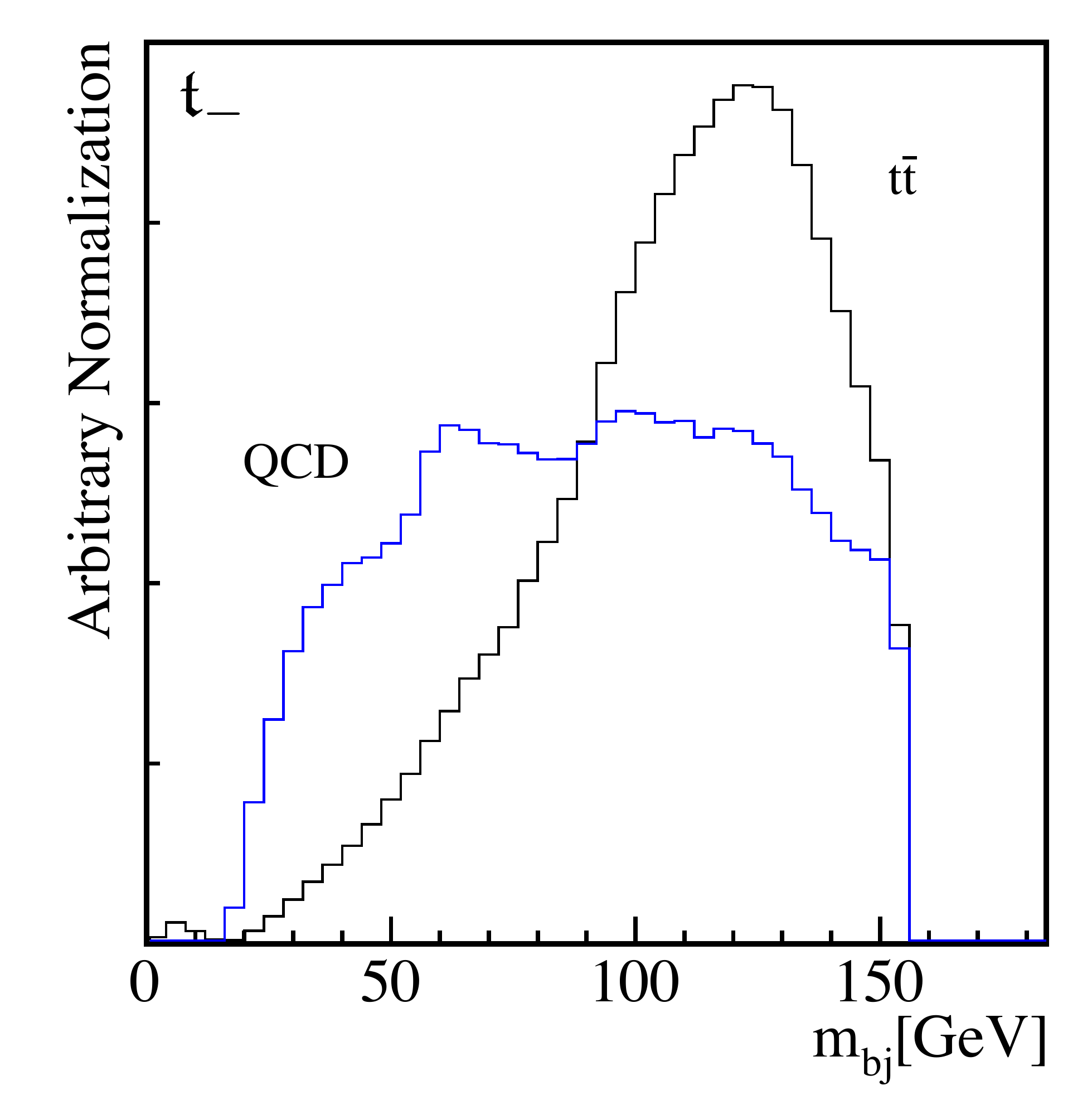}
\caption{Invariant mass distributions of the $b$ quark and the harder
  $W$ decay jet.  Left: parton level with (solid) and without (dotted)
  the requirement $p_{T,j2}< 25~\gev < p_{T,b},p_{T,j1}$. 
  Right: $m_{bj}$ distributions for \xfail buckets. Black lines show $t_h\bar{t}_h$+jets, blue
  lines QCD jets events.}
\label{fig:bj}
\end{figure}

In Section~\ref{sec:simple} we found that it is not rare for the softest
top decay jet to fall below the jet $p_T$ threshold.  Attempts to
reconstruct two tops in ($3j$,$3j$)-buckets will then fail.  In 94\%
of these cases the softest of the six top decay partons comes from the
$W$ decay.  Restricted to events where the sixth parton falls below
$25~\gev$ this fraction increases to 98.5\%, \ie~whenever the sixth
parton is missing the surviving two jets are the bottom and the harder
$W$ decay jet.  In Figure~\ref{fig:bj} we first show the invariant mass
of the $b$ and the harder $W$ decay product $m_{bj_1}$ at parton
level.  We see a clear peak and an endpoint $m_{bj_1} < \sqrt{m_t^2 -
  m_W^2} \simeq 155$~GeV~\cite{endpoint}.  For events where the softer
$W$ decay jet falls below the $p_T$ threshold the peak becomes more
pronounced.\bigskip

\begin{figure}[t]
\includegraphics[width=0.3\textwidth]{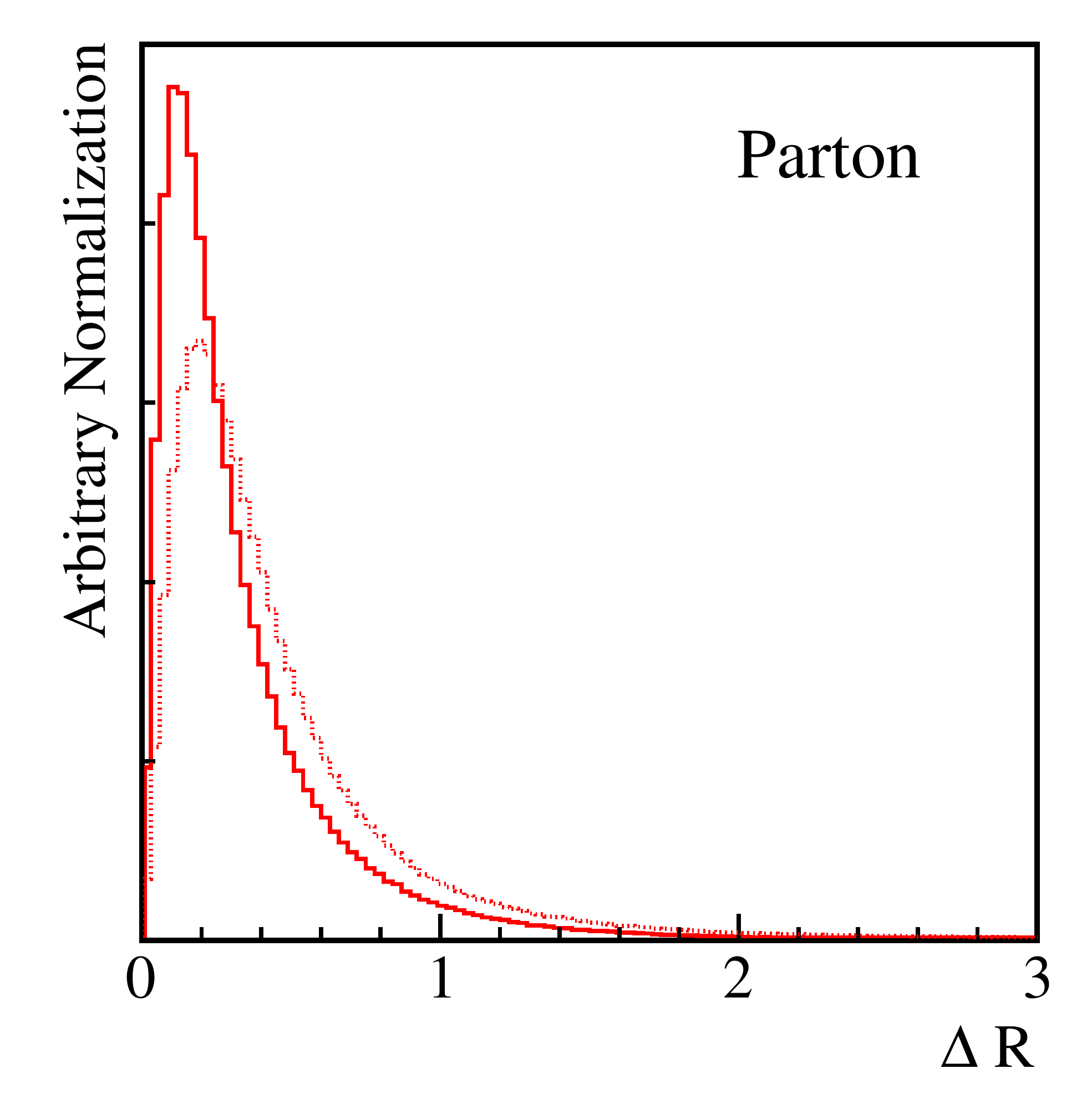} 
\hspace*{0.03\textwidth}
\includegraphics[width=0.3\textwidth]{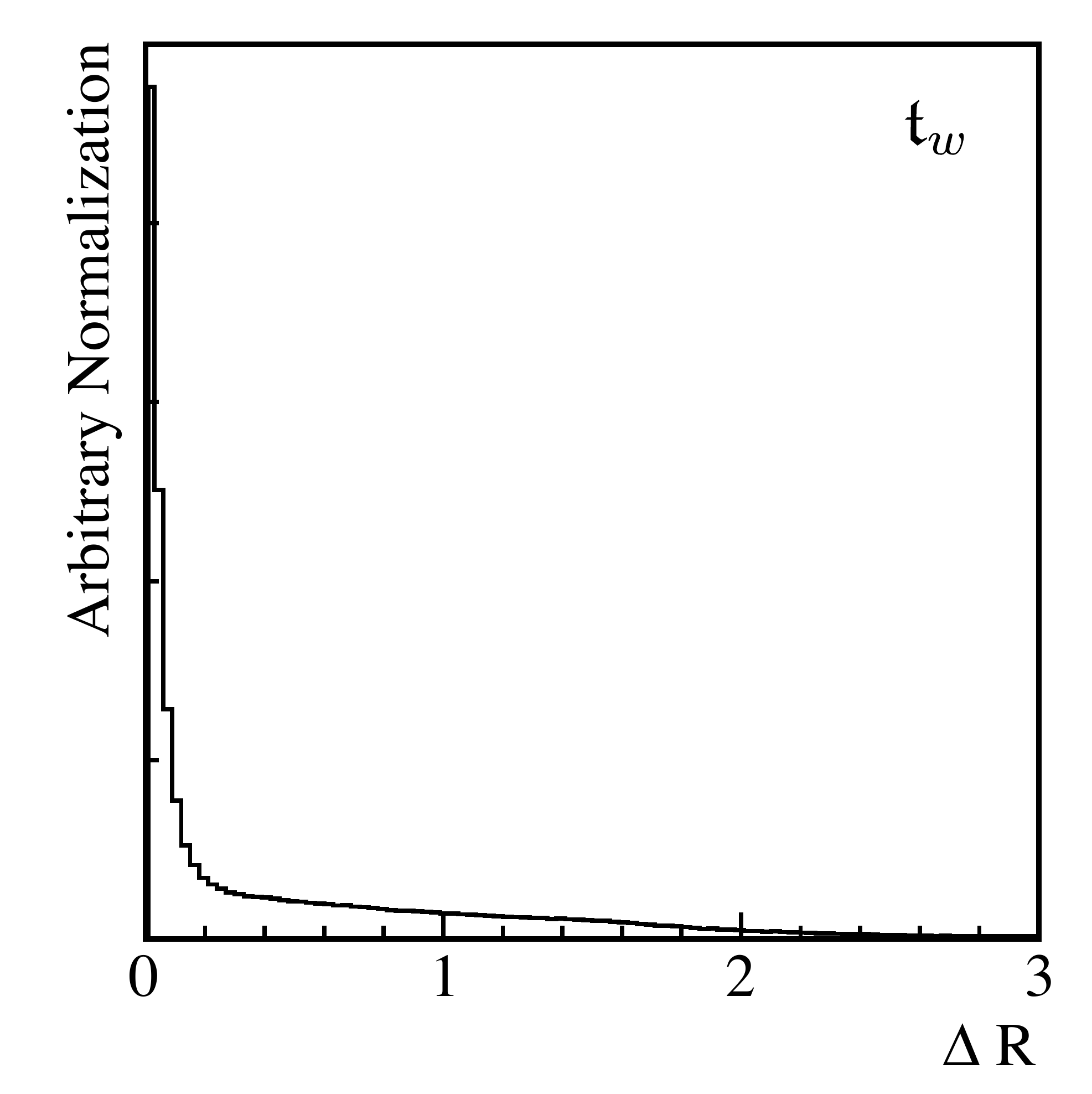}
\hspace*{0.03\textwidth}
\includegraphics[width=0.3\textwidth]{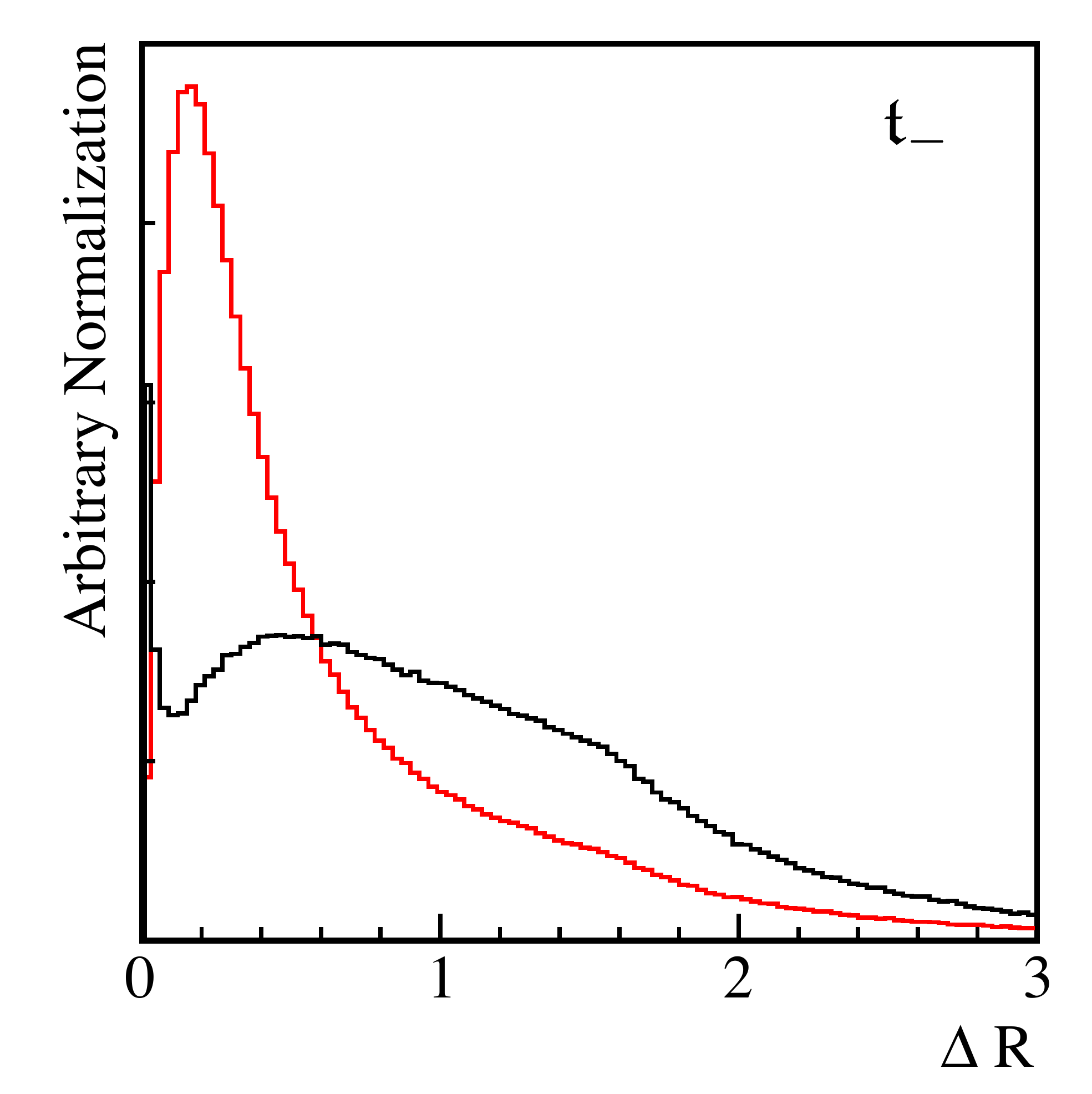} \\
\includegraphics[width=0.3\textwidth]{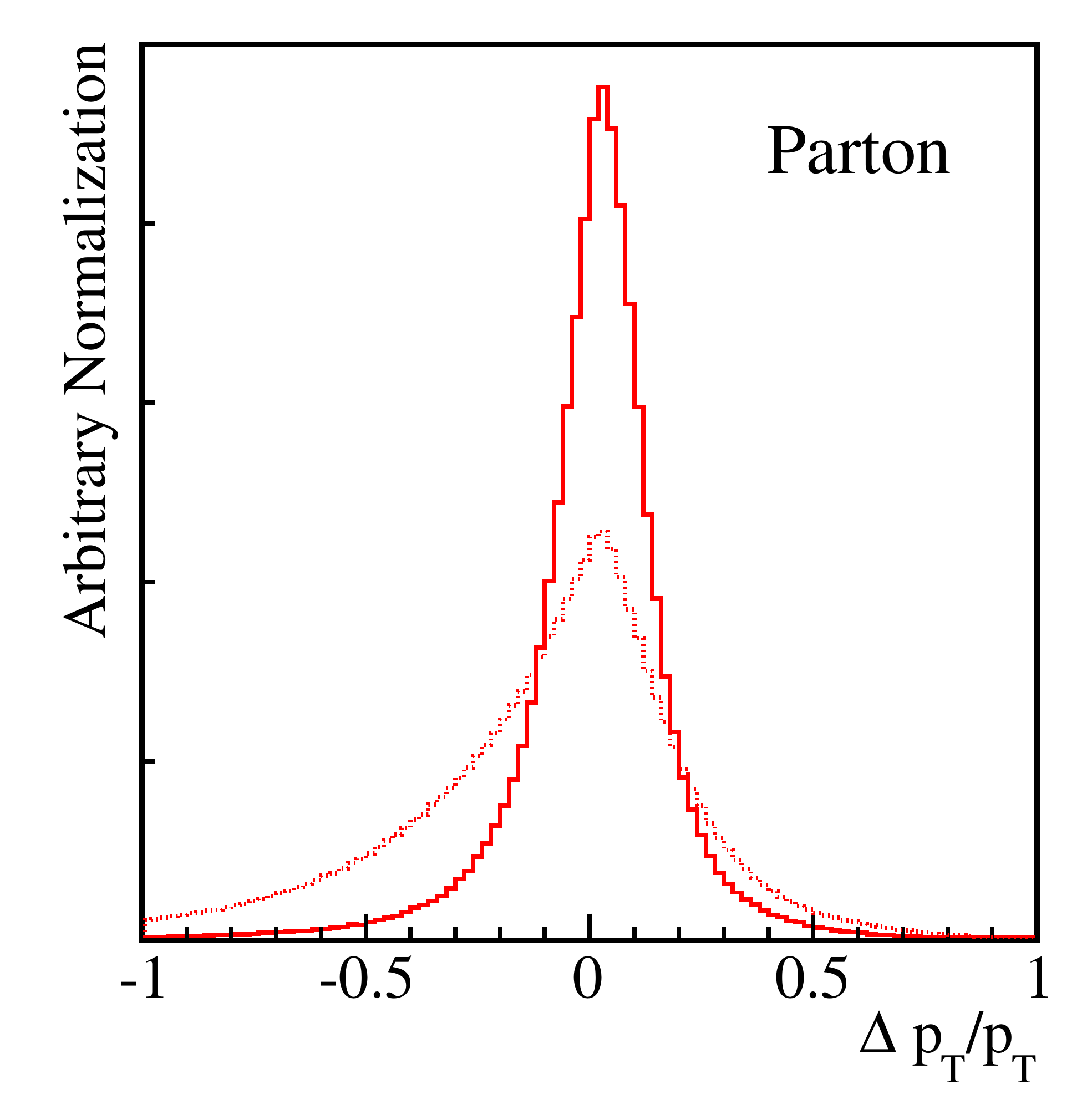}
\hspace*{0.03\textwidth}
\includegraphics[width=0.3\textwidth]{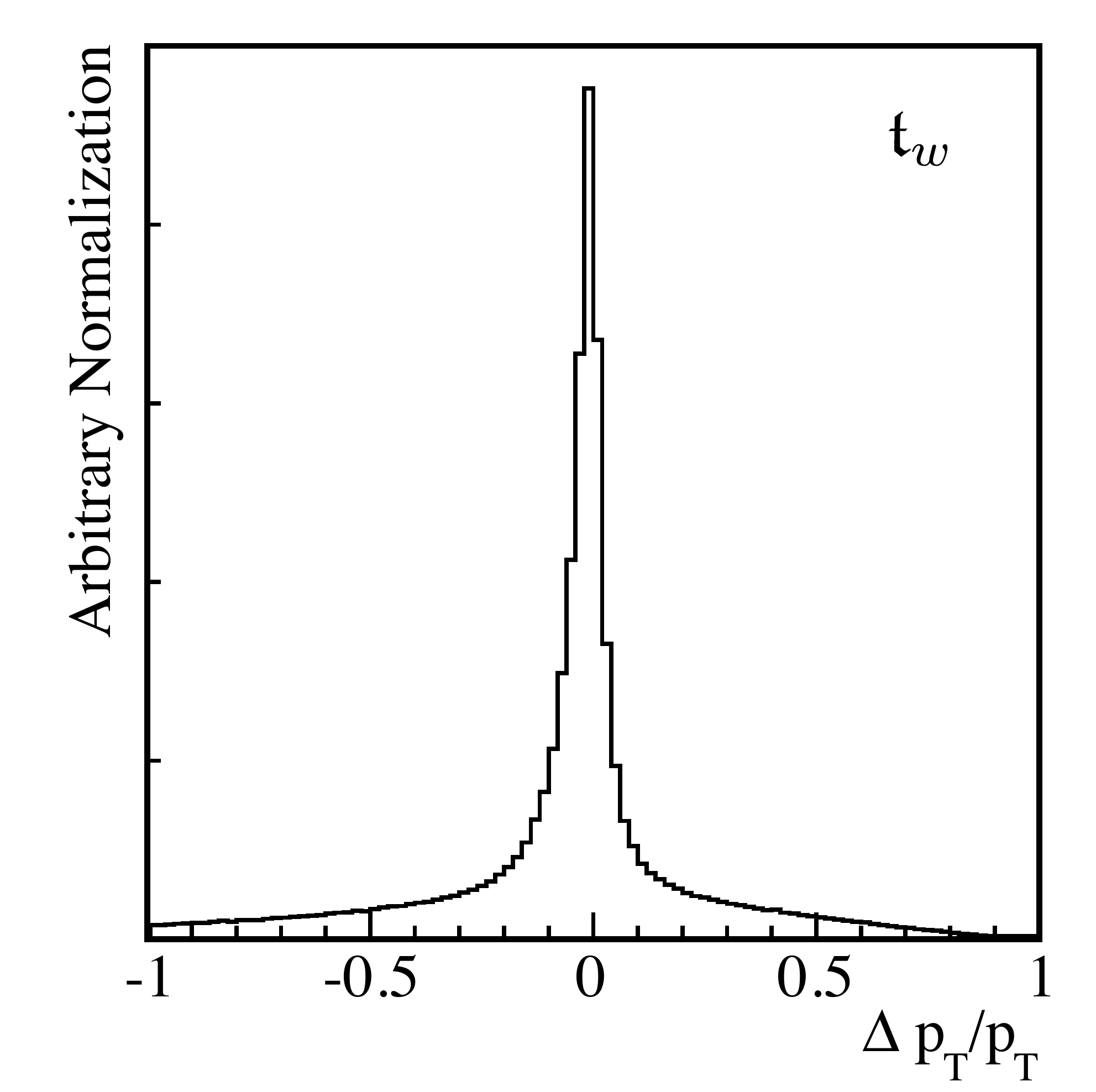}
\hspace*{0.03\textwidth}
\includegraphics[width=0.3\textwidth]{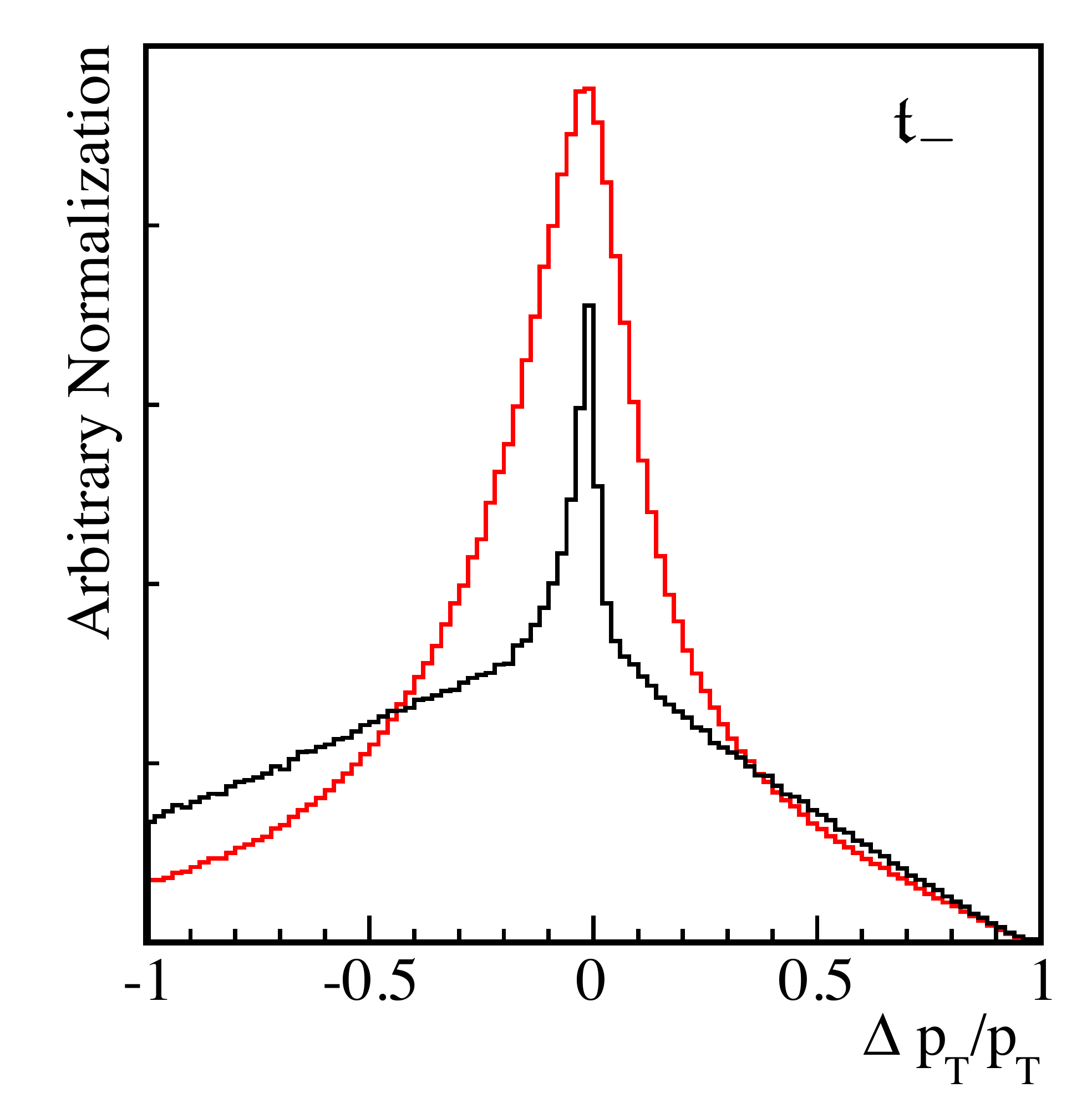}
\caption{Upper Row: $\Delta R$ between top parton and reconstructed
  top from parton level simulation, \xpass\, buckets, and \xfail\, buckets
  (left to right).  
  The dotted line at parton level shows all events, 
  the solid lines only events 
  with $p_T < 25~\gev$ for the 3$^\text{rd}$ top decay product. 
  For reconstructed tops, buckets with a single
  $b$-jet in $B_1$ and $B_2$ without $\Delta^{bj}_B$ minimization are
  are shown in black, \xfail\, buckets after $\Delta^{bj}_B$ minimization in
  red. Lower row: $\Delta p_T/p_T^\text{rec}$ distributions for the same set of
  events.}
\label{fig:topimprovement}
\end{figure}

The question is: can we use the predicted peak in the $m_{bj}$
distribution to identify tops in $2j$-buckets?
If the third missing top decay jet indeed fails the $p_T$ threshold we
expect the top momentum to be close to the $b$/jet momentum. The left
panels of Figure~\ref{fig:topimprovement} show the difference between
the parton-level top momentum and the $b$/jet system in terms of
$(p_{T,bj} - p_{T,t})/p_{T,bj} \equiv \Delta p_T/p_{T,bj}$ and $\Delta
R$. If we assume an $R$ separation of around 0.5 as a quality measure the
result looks promising. Similarly, we should be able to reconstruct
the transverse momentum of the top at least at the 20\% level without
including the softest $W$ decay jet. In comparing our bucket method to top taggers, it should
be emphasized that for the bucket method we allow for a missing $W$
decay jet rather than replacing the lighter $W$ decay jet by a QCD
jet~\cite{HEP3}.\bigskip


For $2j$-buckets, which we know do not include all three top decay products, 
we replace the distance measure of Eq.~\eqref{eq:delta} with a
similar measure inspired by the distribution of $m_{bj}$,
\begin{equation}
\Delta^{bj}_B = \left\{
\begin{array}{ll}
|m_B - 145~\gev|  & \qquad \text{if} \ m_B \le 155~\gev\\
\infty & \qquad  \text{else}\\
\end{array}
\right. \; .
\label{eq:Bjwindow}
\end{equation}
The peak value of 145~GeV is read off Figure~\ref{fig:bj} and should
eventually be tuned to data.

Because $3j$-buckets already reconstruct the top momentum we keep
them.  For top buckets in the (\xpass,\xfail), (\xfail,\xpass), and
(\xfail,\xfail) categories which do not contain a $W$ candidate we
re-assign jets replacing Eq.~\eqref{eq:delta} with the new distance
measure. In addition, we need to remove the top mass selection cut
Eq.~\eqref{eq:mBcut}. This way combinations of $b$ quarks and jets
which do not fall into the window of Eq.~\eqref{eq:Bjwindow} are kept.
The new reconstruction algorithm reads
\begin{itemize}
\item (\xpass,\xpass): keep these buckets as is,
\item (\xpass,\xfail) or (\xfail,\xpass): reconstruct the failed bucket using
  all non-\xpass\, jets, minimizing $\Delta^{bj}_B$,
\item (\xfail,\xfail): use all jets to minimize $\Delta^{bj}_{B_1} +
  \Delta^{bj}_{B_2}$.
\end{itemize}
Note that for reconstructing $b$/jet-buckets we use jets both from the
\xfail\, bucket and from the ISR bucket.\bigskip

Comparing to the original algorithm we have adapted the metric for
assigning jet for top buckets in the \xfail\, category. What remains is to
replace the top mass window in Eq.~\eqref{eq:Bjwindow} with
appropriate $b$/jet values. In the right panel of
Figure~\ref{fig:bj} we show the $b$/jet bucket mass distributions
$m_{bj}$ for signal and background.  For the signal they agree well
with the expectation from the left panel of Figure~\ref{fig:bj}.  For
a top candidate we require at least one $b$/jet pair satisfying
\begin{equation}
75~\gev < m_{bj} < 155~\gev \; .
\label{eq:mbj}
\end{equation}
We show the signal and background efficiencies of this new
reconstruction algorithm in Table~\ref{tab:bjbucket_all}, along with
the percentage of correct top reconstruction.  The numbers need to be
compared to Table~\ref{tab:nj_bimposed_summary}. First, we see that
the number of events which contain valid top buckets in the correct
mass window, albeit including one $2j$-bucket, has significantly
increased. In the (\xpass,\xfail) category roughly half of all events
reconstruct both tops well, in spite of missing one of the six decay
jets. The number of (\xfail,\xpass) events passing this reconstruction algorithm drops
significantly when compared to Table~\ref{tab:nj_bimposed_summary}.
Most of these events contain one $b$-jet and one non-$b$-tagged
jet in $B_1$. However, the $b$-jet in this category of events is typically
a merger of a $b$ and the third jet from the top decay. Thus, while
the bucket itself has an invariant mass near the top, it contains neither
a $W$ candidate nor a $b$-jet that can be combined with another
jet in the event to pass the selection criteria in Eq.~\eqref{eq:mbj}. 
Even in the (\xfail,\xfail) category where neither of the two buckets include a
reconstructed $W$ candidate the fraction of well reconstructed top
pairs reaches almost 40\%.\bigskip

\begin{table}[t]
\begin{tabular}{l||r|rrr||rr||r}
\hline
 &  $t_h\bar{t}_h$+jets [fb] & $R_1, R_2<0.5$ &$R_{1}<0.5$ & $R_{2}<0.5$ &QCD [fb] & $W$+jets [fb] & $S/B_\text{QCD}$  \\
\hline
5 jets, 2$b$-tag &21590& &&& 16072 &    109.6& 1.4  \\
\hline
(\xpass,\xpass)  & 2750 & 68.9\% & 9.3\% & 7.5\% & 126.2 & 1.181 & 21.8 \\
\hline
(\xpass,\xfail)   & 7465 & 49.0\% & 17.8\% & 10.3\% & 2145 & 15.78 & 3.5 \\
(\xfail,\xpass)& 997 & 29.5\% & 19.7\% & 16.9\% & 160.2 & 1.42 & 6.2\\
(\xfail,\xfail)  & 3979 & 38.7\% & 17.0\% & 15.1\% & 2575 & 17.49 & 1.6 \\
\hline
\end{tabular}
\caption{Number of events reconstructed using the $b$/jet-buckets for
  (\xpass,\xfail), (\xfail,\xpass) and (\xfail,\xfail) events. The numbers for
  (\xpass,\xpass) events are unchanged from
  Table~\ref{tab:nj_bimposed_summary}.}
\label{tab:bjbucket_all}
\end{table}

To study the quality of the top reconstruction in more detail we show
the difference between the bucket momentum and the parton level top
momentum in terms of $\Delta R$ and $\Delta p_T/p_T$ in the right two
panels of Figure~\ref{fig:topimprovement}. The buckets constructed 
around $b$-jets are shown in black. The
results of replacing the \xfail\, buckets using the $b$/jet algorithm are
shown in red. In this case we see a narrow peak at zero which corresponds
to complete top momentum reconstruction where we fail to find a
$W$ candidate due to overlapping jets. Such events - which are in the minority
 - often fail to pass the reconstruction using the $\Delta_B^{bj}$ metric. As
 a result, the narrow peak at zero is not present in this second reconstruction method.


For \xfail buckets the $b$/jet algorithm
consistently reconstructs the top direction significantly better than using the
original method.
In contrast, changing \xpass buckets to the $b$/jet-bucket does not
improve the momentum reconstruction. 
We checked $b$/jet-momentum provides
better top momentum reconstruction than only using the bottom momentum.



\subsection*{${\mathbf p_T}$ dependent efficiencies}

Until now we have focused on identifying and reconstructing pairs of
hadronically decaying top quarks from the complete signal sample. The
results shown in Table~\ref{tab:bjbucket_all} indicate that the
efficiency as well as the background rejection of our algorithm allows
for a systematic study of hadronic top pairs. However, the fraction of
events with not-quite-perfect reconstruction of the top directions
($R_i > 0.5$ for $i=1,2$), is somewhat worrisome. From top
tagging we know that a certain fraction of relatively poorly
reconstructed tops cannot be avoided~\cite{HEP3}, but that fraction
should be small. What we need is a self-consistency requirement --- or
QMM\footnote{We use this opportunity to, for the first time, introduce
  `quality management milestones' (QMM) in a research paper.} --- similar
to only accepting reconstructed tops with $p_{T,t} > 200$~GeV in a top
tagger~\cite{HEP1}.\bigskip

\begin{figure}[t]
\includegraphics[width=0.30\textwidth]{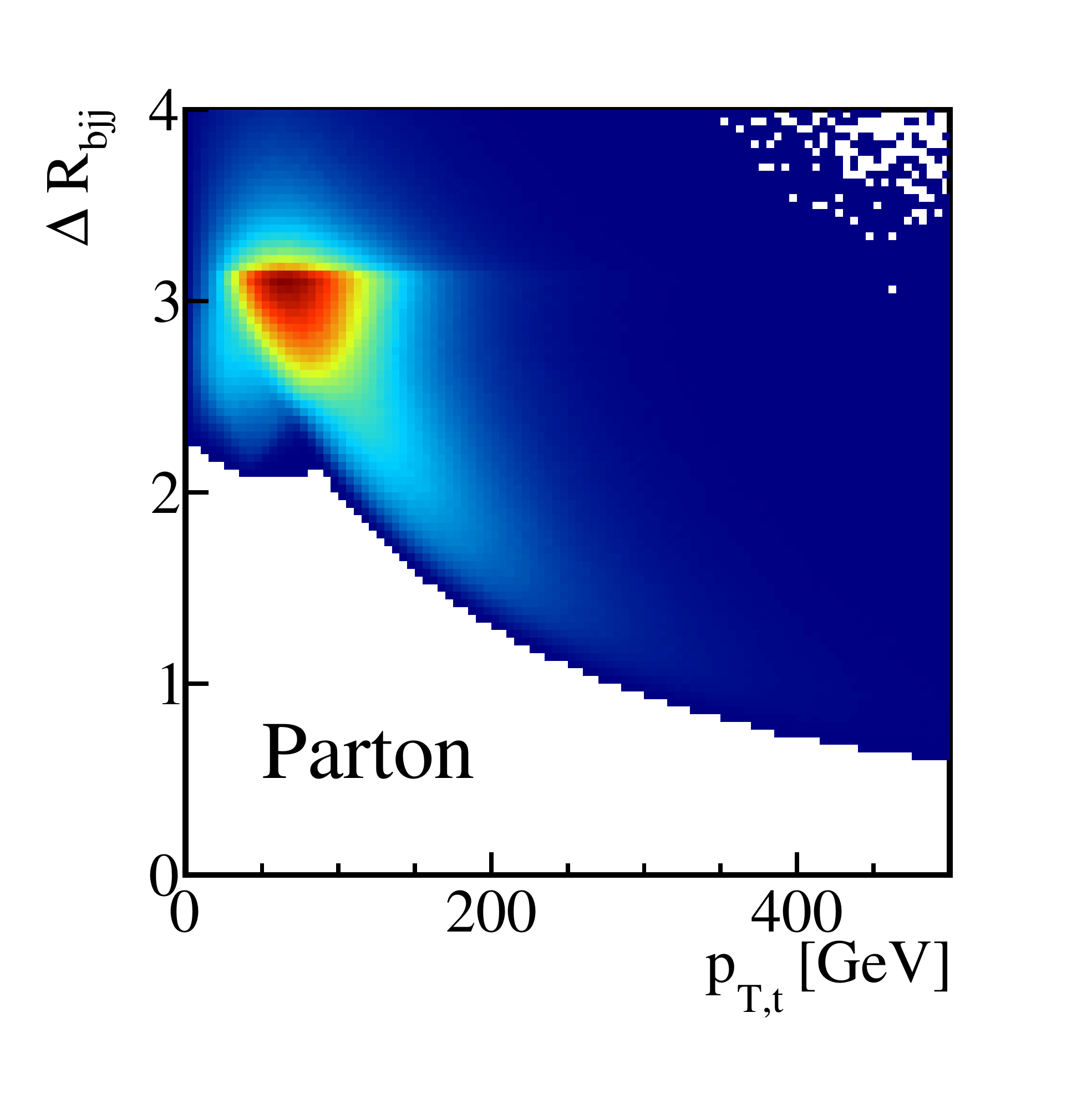}
\includegraphics[width=0.30\textwidth]{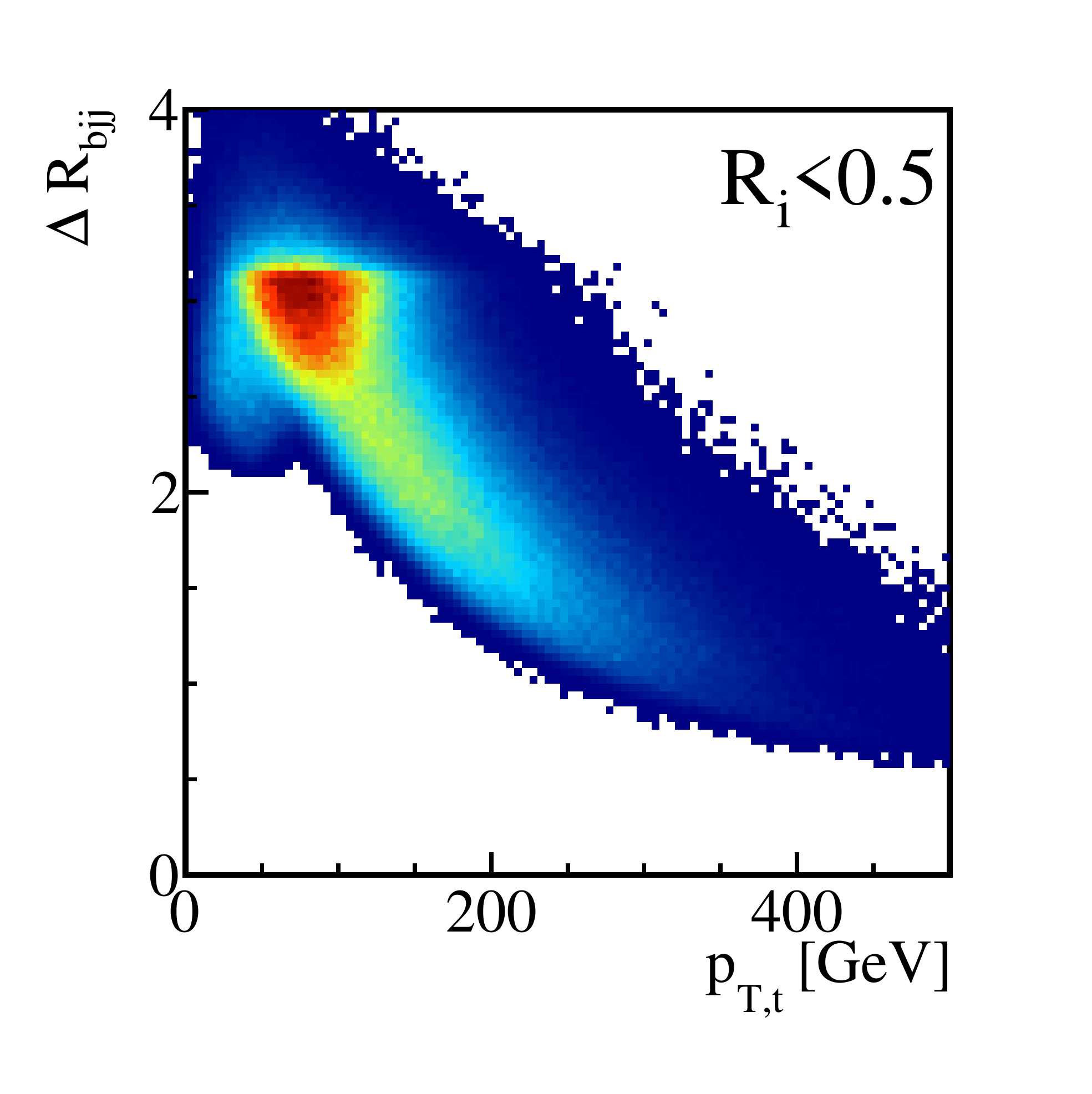}
\includegraphics[width=0.30\textwidth]{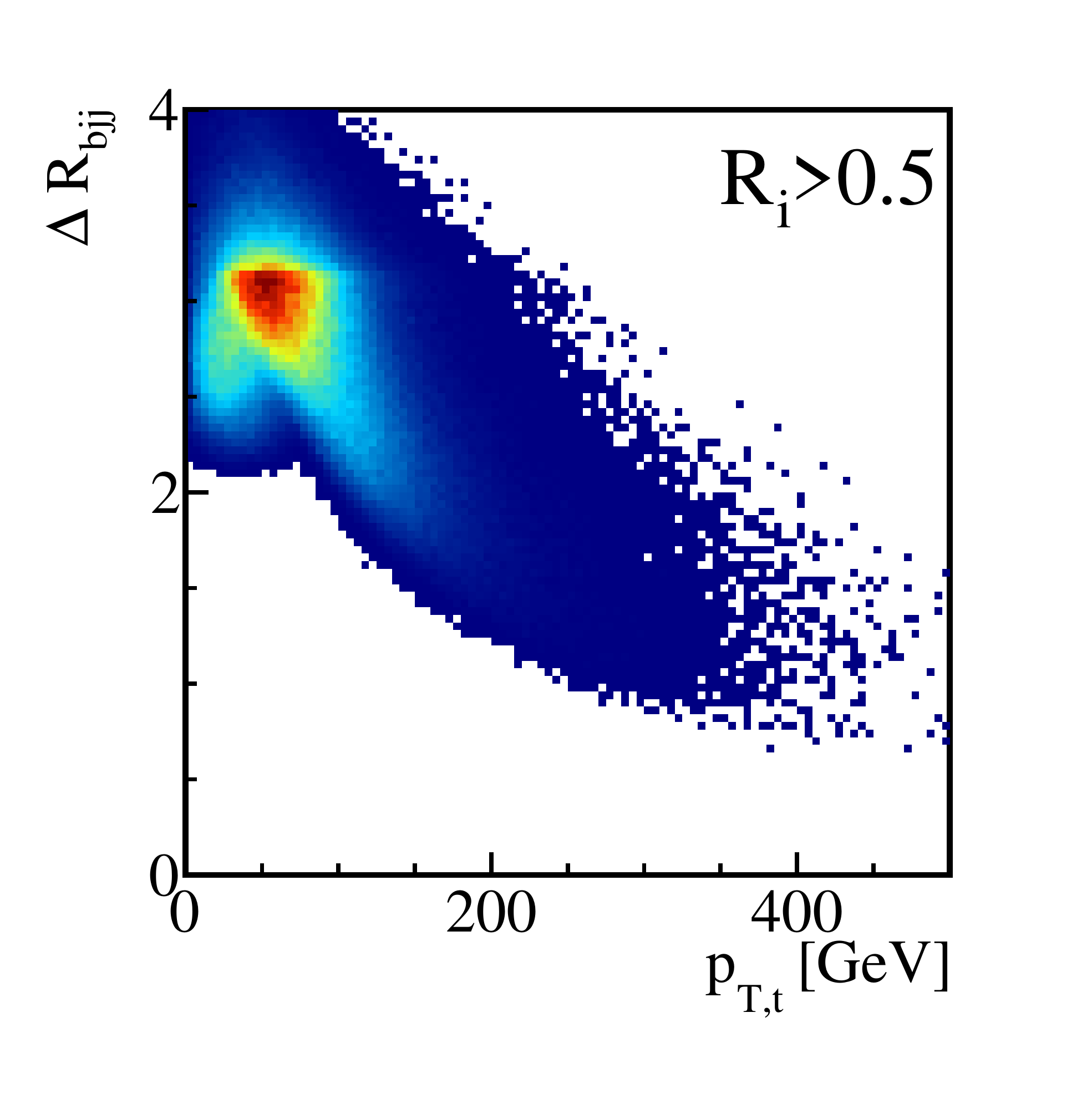}
\\
\includegraphics[width=0.30\textwidth]{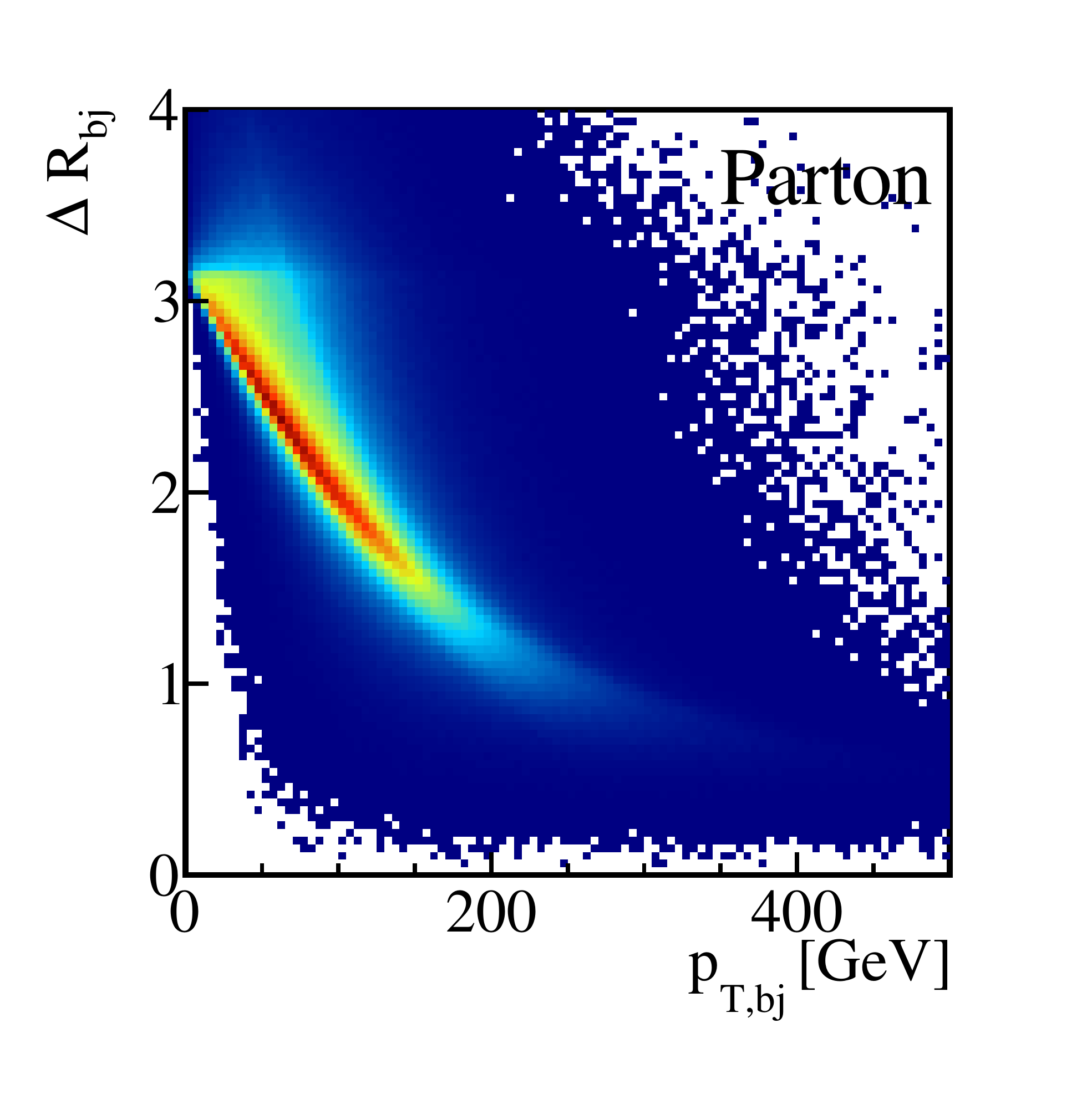}
\includegraphics[width=0.30\textwidth]{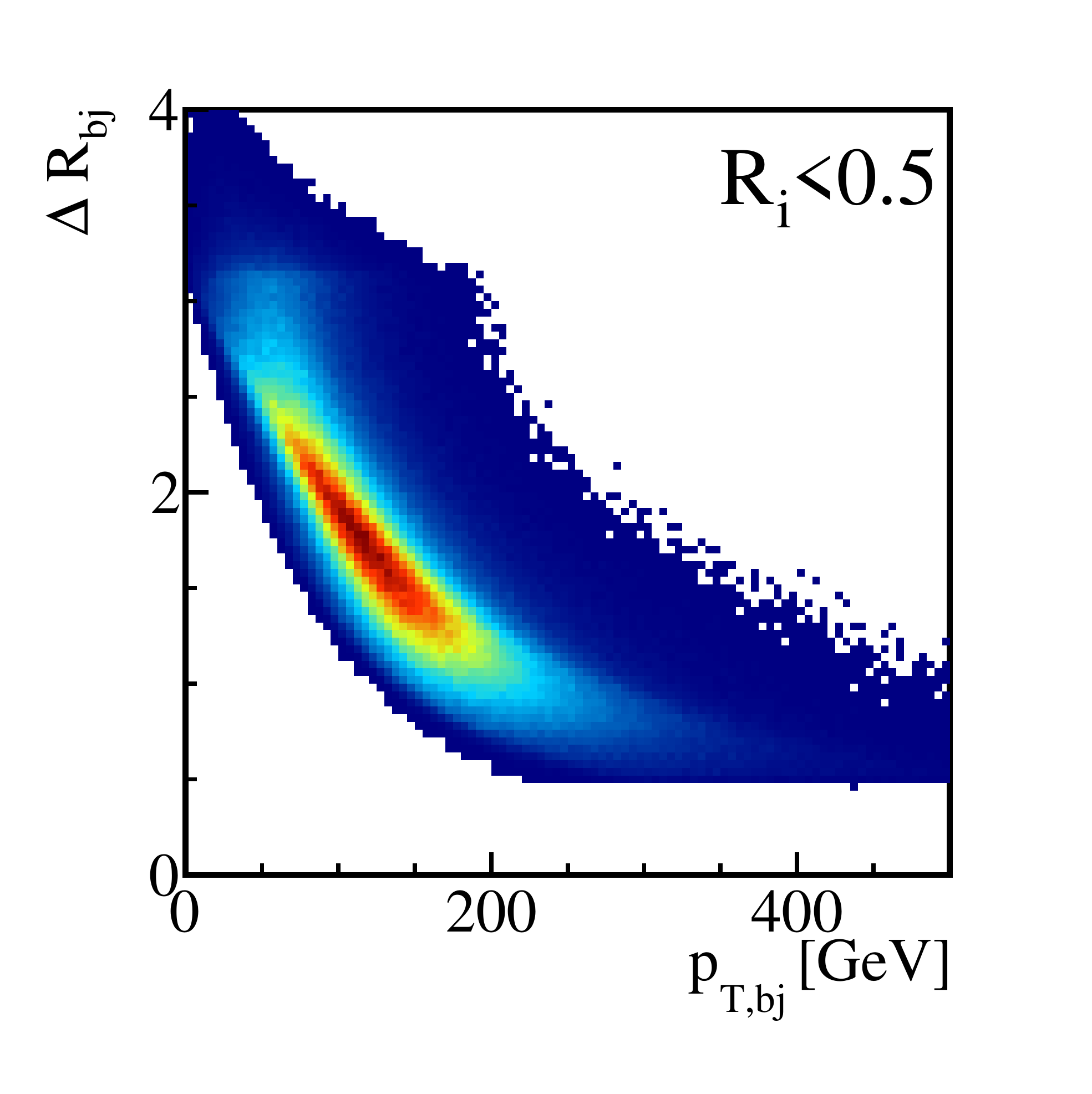}
\includegraphics[width=0.30\textwidth]{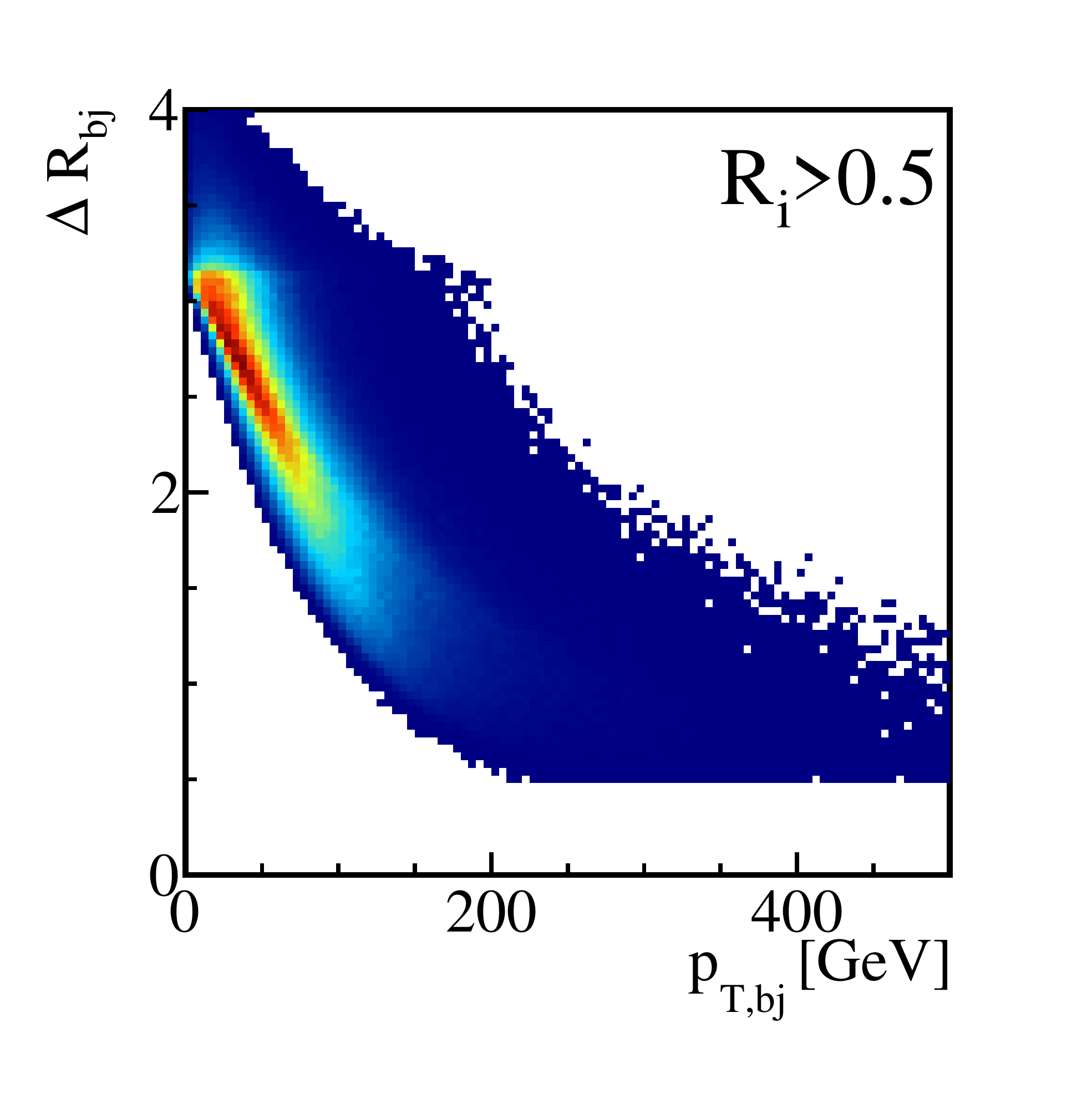}
\caption{Correlation between $p_T$ and the $\Delta R_{bj}$ (for $(2j)$-buckets) or $\Delta R_{bjj}$ (for $(3j)$-buckets) for
  reconstructed hadronic tops. From left to right: parton level
  simulation, buckets correctly reconstructing the top direction ($R_i
  < 0.5$), buckets not reconstructing the top direction $(R_i >
  0.5)$. The upper row shows \xpass\, buckets, the lower row \xfail\,
  buckets.}
\label{fig:consistency}
\end{figure}

Once we identify a top buckets we can use two observables to define such a
QMM: the top momentum and the geometric size of
the hadronic top decay. The latter is defined differently for \xpass\,
buckets and \xfail\, buckets. In the first case we have access to all
pair-wise $\Delta R$ distances between the three top decay
products. We define $R_{bjj}$ as the maximum of the three $\Delta R$
separations of the top decay products. For \xfail\, buckets we only
have one distance, namely $R_{bj}$ between the bottom and the hardest
light-flavor jet. 

In Figure~\ref{fig:consistency} we show the correlation between
these two observables, first for parton level simulations in the left
column. For both kinds of buckets we see a clear correlation, with the main
difference being that most \xpass\; buckets have relatively low transverse
momenta. For \xfail\, buckets, which require the softest top decay jet
to fall below $p_{T,j} = 25$~GeV, the distribution extends to larger
transverse momenta where the initial boost of the top can compensate
or the decay momentum of the softest jet.

The second column shows the reconstructed observables for \xpass\, and
\xfail\, buckets, requiring that the buckets reconstruct the parton level top
direction within $R < 0.5$. The correlation between size and
transverse momentum is the same as expected from simulation. However,
we clearly see that either large transverse momenta, $p_{T,t} \gtrsim
100$~GeV, or small sizes, $\Delta R_{bj(j)} \lesssim 2.5$, are preferred. This is particularly true for
\xfail\, buckets. The reason for this is that a slight boost of the top quarks
generates a geometric separation of the transverse back-to-back tops
and the forward ISR jets.  Combinations of jets from different buckets
are now separated in their typical transverse mass values. This gives
us a handle on combinatorics and improves the top reconstruction even
in the case where one of the top decay products is missing.

Conversely, buckets passing as tops but giving a poor directional
reconstruction reside at low transverse momenta and large size, as can
be seen in the third column of Figure~\ref{fig:consistency}. To veto
these buckets we have a choice of criteria in the two-dimensional
$R_{bj(j)}$ vs. $p_{T,t}$ plane. We choose the condition
\begin{alignat}{5}
p_{T,t}^\text{rec} > 100~\gev
\label{eq:consistency}
\end{alignat}
at the level of the buckets to increase the fraction of well
reconstructed or matched top quarks in both bucket categories. This
choice results in the highest efficiency of well-reconstructed tops in
both \xpass\, and \xfail\, buckets. 
Alternative conditions in terms of
$R_{bj(j)}$ or in the two-dimensional planes shown in
Figure~\ref{fig:consistency} could replace Eq.~\eqref{eq:consistency} in
specific analyses.
For example, a stricter cut will result  in higher purity of well-reconstructed tops.
\bigskip

\begin{figure}[t]
\includegraphics[width=0.32\textwidth]{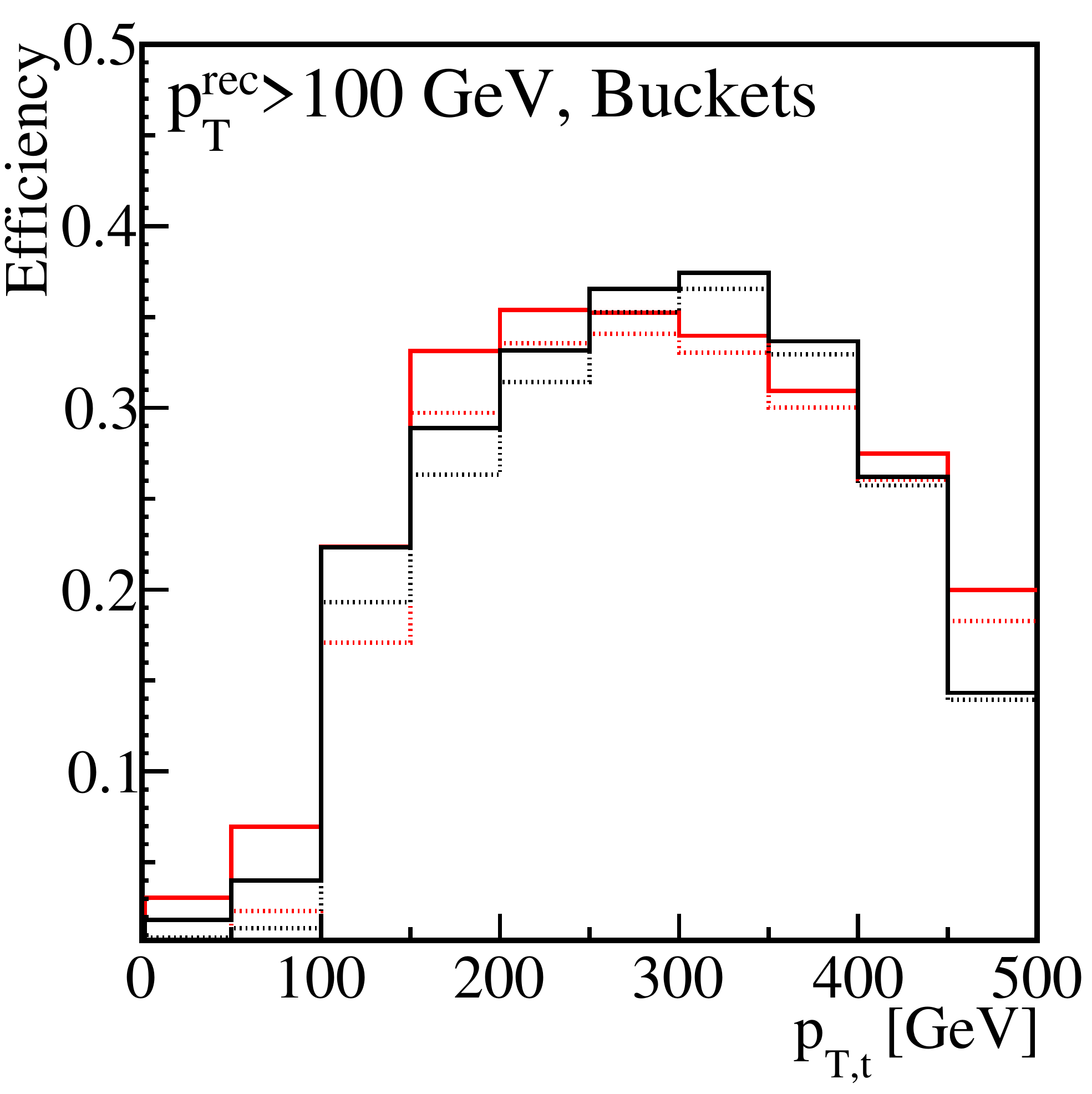}
\includegraphics[width=0.32\textwidth]{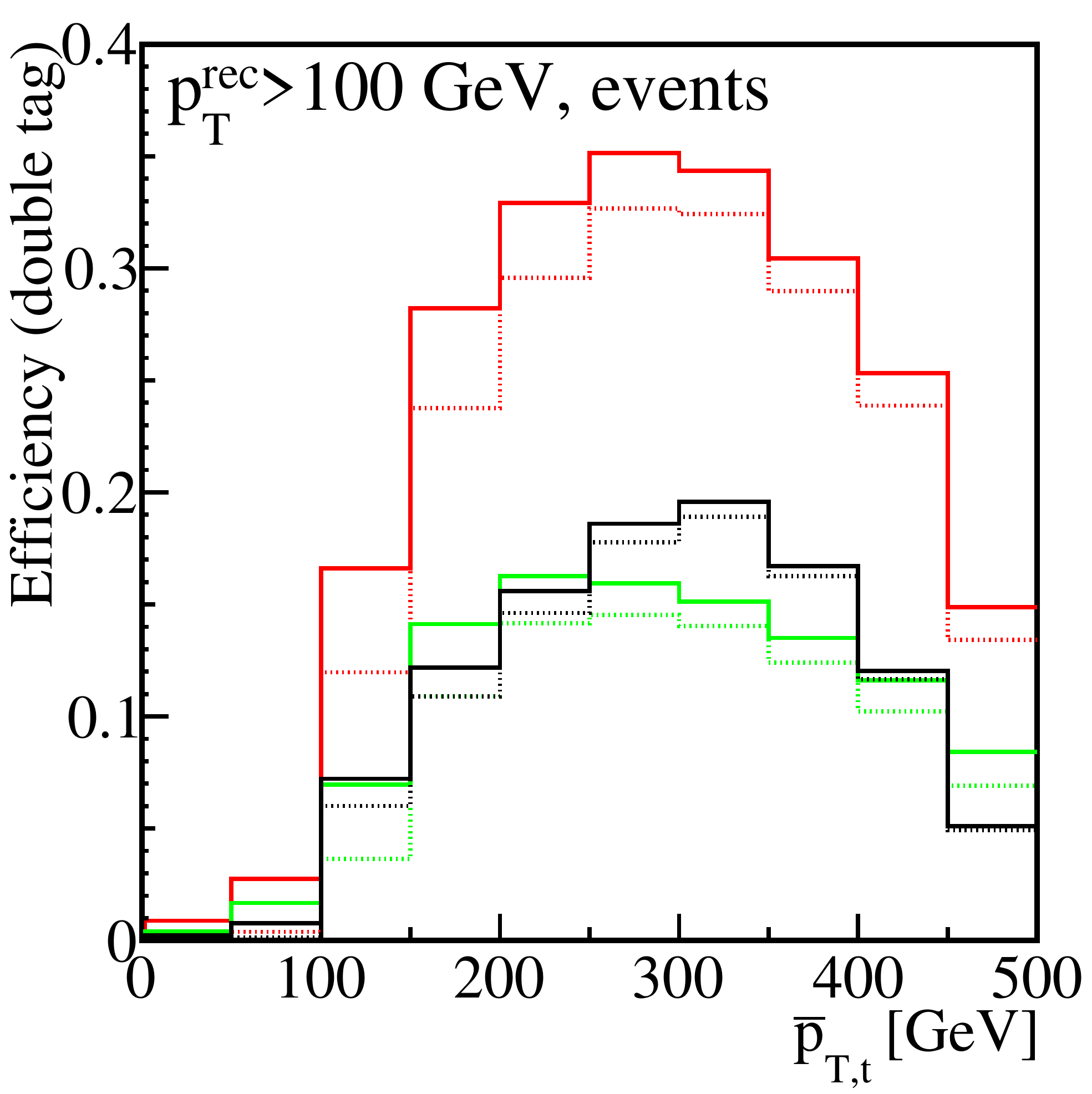}
\includegraphics[width=0.32\textwidth]{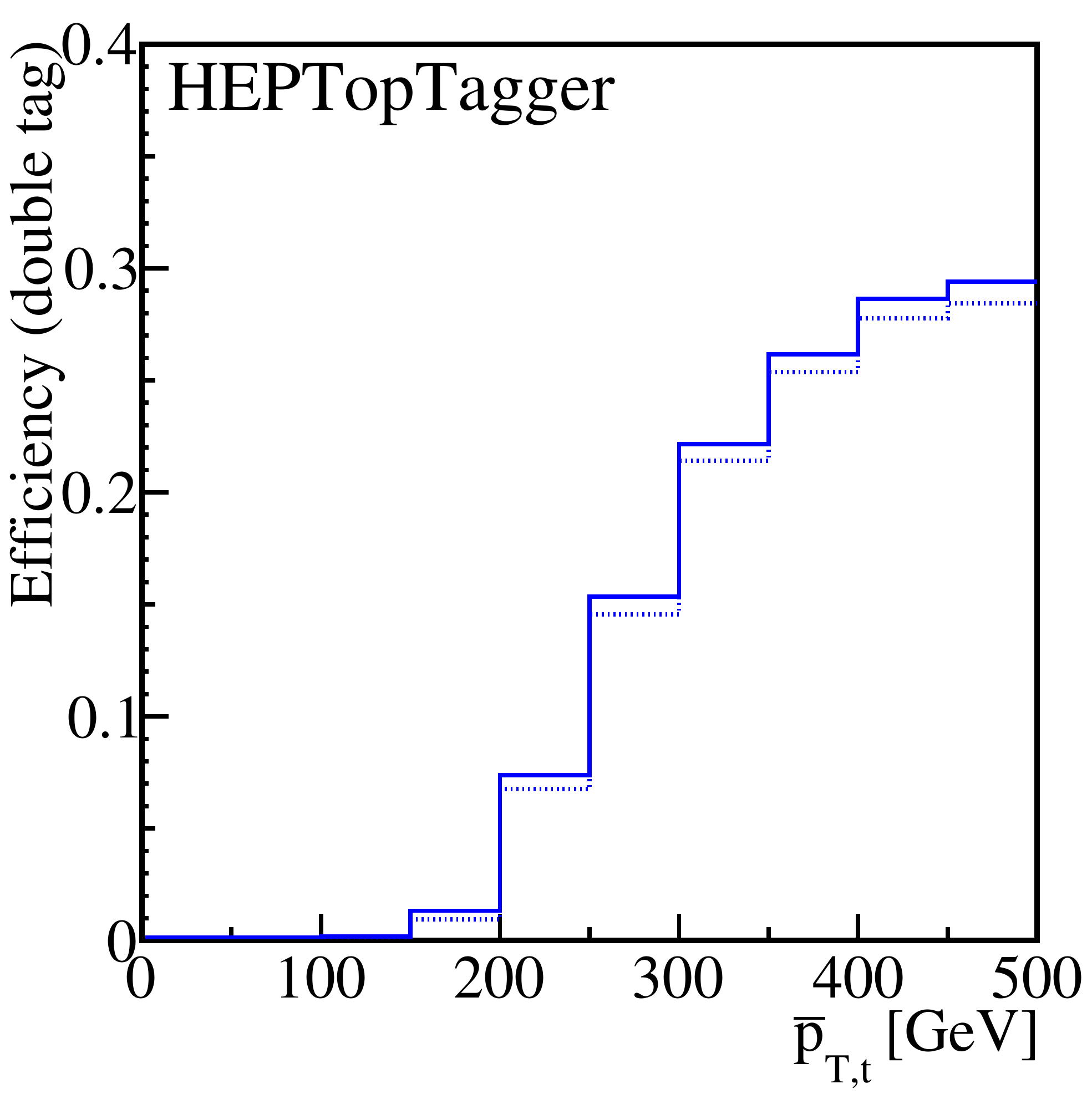}
\caption{Left: efficiency for a single bucket tag as a function of the
  true transverse momentum of the top, shown for \xpass\, buckets
  (black) and \xfail\, buckets (red). Dashed lines indicate events
  matched to a top quark within $R_i < 0.5$.  Center: efficiencies for
  two bucket tags as a function of the average true $p_{T,t}$. We
  show (\xpass,\xpass) events in black, (\xpass,\xfail)/(\xfail,\xpass) in red,
  and (\xfail,\xfail) event in green. Dashed lines again indicate reconstructed tops matched to parton level tops.
  Right: corresponding efficiency for two
  {\sc HEPTopTagger}~\cite{tth,HEP1} tags.  In all cases the last bin
  includes all events above 450~GeV.}
\label{fig:eff}
\end{figure}

To illustrate the power of the bucket algorithm we compute the efficiency for
reconstructing a single top as well as a top pair as a function of
the transverse momenta of the tops. The left panel of Figure~\ref{fig:eff} shows the efficiency for
a bucket tag as a function of the true transverse momentum of
the top. The baseline is all fully hadronic $t \bar{t}$ events in the
Standard Model, with five or more jets and two $b$-tags.  A possible
mis-measurement of $p_{T,t}$ in particular at low transverse momenta
explains the tail of events below the apparent consistency criterion
$p_{T,t}^\text{rec} > 100$~GeV. We see that the tagging efficiency
increases rapidly right at threshold. Above $p_{T,t} = 150$~GeV more
than 90\% of the tagged top quarks can be matched to a true top within
$R_i < 0.5$. For $p_{T,t}=100 - 150$~GeV  about 80\% can be so matched.
For \xpass\, and \xfail\, buckets the number of unmatched
tops becomes negligible above 250~GeV. Adding \xpass\, and \xfail\, buckets, 
the total efficiency of our algorithm is 60-70\% for $150< p_{T,t} < 350$~GeV.

In the central panel of Figure~\ref{fig:eff} we show the tagging
efficiency for two top quarks as a function of the average true
transverse momentum $\bar{p}_T=(p_{T,t1}+p_{T,t2})/2$.  The total
efficiency is split between (\xpass,\xpass) events (black),
(\xpass,\xfail) or (\xfail,\xpass) events (red), and (\xfail,\xfail)
events (green). For each of these categories we also show the well
reconstructed tops only. As expected, the (\xpass,\xpass) events are
reconstructed with an encouragingly high efficiency and essentially
negligible number of non-matched tops. For the other two categories the fraction
of unmatched tops is slightly larger, but well under control. 

Also note that the efficiency for (\xpass,\xpass) events is slightly higher than
the square of the single bucket \xpass\, efficiency. This is because, once one 
top in an event is reconstructed, the second top becomes easier to find, due to 
combinatorial factors.  Similar correlations occur in the (\xpass,\xfail), (\xfail,\xpass) and
(\xfail,\xfail) categories. The total double top tag efficiency for $\bar{p}_{T,t}=150-350$~GeV
is close to the single tag efficiency: 55-70\%. As we always search for two tops
(otherwise we regard the event as un-reconstructed), the total double tag efficiency 
and total bucket tag efficiency must be closely related, as long as the individual 
$p_{T,t}$ and averaged $\bar{p}_{T,t}$ distributions are similar. We should note that some of the unmatched tops may still be correct tags
as QCD effects will change the direction of the true top as
compared to the top decay products at parton and particle
level.\bigskip

The resulting cross sections of reconstructed tops with the consistency selection cut Eq.~\eqref{eq:consistency}
are summarized in Table~\ref{tab:number_pt100}. The total double top tag 
efficiency for the $t_h\bar{t}_h$+jets sample with five jets of which two are $b$-tagged is 28\%.
The mis-tagging efficiency of finding two valid top
buckets in the pure QCD events (five jet, two mis-tagged as $b$-jets) is of the order of 5\%.

Unlike for a typical top tagger, illustrated in the right panel, the
efficiency of the buckets does not reach a plateau at large transverse
momentum. Once the top decay jets start merging at the scale of the
C/A jet size the method will fail, so for example $R_\text{C/A}=0.5$
leads to a drop above $p_{T,t} \gtrsim m_t/R \sim 350$~GeV. Towards
smaller top momenta the requirement Eq.~\eqref{eq:consistency} limits
the efficiency by removing poorly reconstructed tops due to
combinatorics. By construction, the bucket method targets the
intermediate regime $150~\gev < p_{T,t} < 350~\gev$ where it should
serve as a very useful tool in Higgs searches as well as new physics
searches.

\begin{table}[t]
\centering
\begin{tabular}{l||r|r|r|r||rr|r}
\hline
 &  $t_h\bar{t}_h$+jets [fb] & $R_{1}, R_2<0.5$ &$R_{1}<0.5$ & $R_{2}<0.5$ &QCD [fb] & $W$+jets [fb] & $S/B_\text{QCD}$  \\
\hline
5 jets, 2$b$-tag &21590& &&& 16072 &    109.6& 1.36  \\
\hline
(\xpass,\xpass), $p_T^\text{rec}>100$~GeV & 1417 & 86.4\% & 5.4\% & 4.1\% & 27.1 & 0.34 & 52.3 \\
(\xpass,\xfail), $p_T^\text{rec}>100$~GeV  & 2875 & 80.1\% & 7.4\% & 6.2\% & 308.3 & 3.44 & 9.3 \\
(\xfail,\xpass), $p_T^\text{rec}>100$~GeV  & 309.1 & 60.2\% & 15.4\% & 13.3\% & 26.6 & 0.33 & 11.6 \\
(\xfail,\xfail), $p_T^\text{rec}>100$~GeV  & 1507 & 68.5\% & 11.1\% & 11.8\% & 417.2 & 4.69 & 3.6 \\
\hline
total, $p_T^\text{rec}>100$~GeV   & 6109 & 77.7\% & 8.2\% & 7.5\% & 779.2 & 8.81 & 7.8 \\
\hline
\end{tabular}
\caption{Number of events reconstructed using the $b$/jet-buckets for
  (\xpass,\xfail), (\xfail,\xpass) and (\xfail,\xfail) events with $p_{T,t}^\text{rec}>100$~GeV cut.}
\label{tab:number_pt100}
\end{table}

\section{Stops from buckets}
\label{sec:stop}

As a demonstration of our algorithm for top reconstruction, we apply
it to scalar top searches. Searches for supersymmetry, or general, top
partners are becoming more and more central in ATLAS and CMS. They
constrain the allowed stop masses to $m_{\tilde{t}} \gtrsim
600$~GeV~\cite{stops_ex}. Theoretically, many analysis strategies have
been suggested, covering the semileptonic decay
channel~\cite{stops_semilep}, the hadronic decay
channel~\cite{stops_had}, or dedicated {\sc HEPTopTagger} studies in
each of these channels~\cite{stops_hep}.

\begin{figure}[t]
\includegraphics[width=0.4\textwidth]{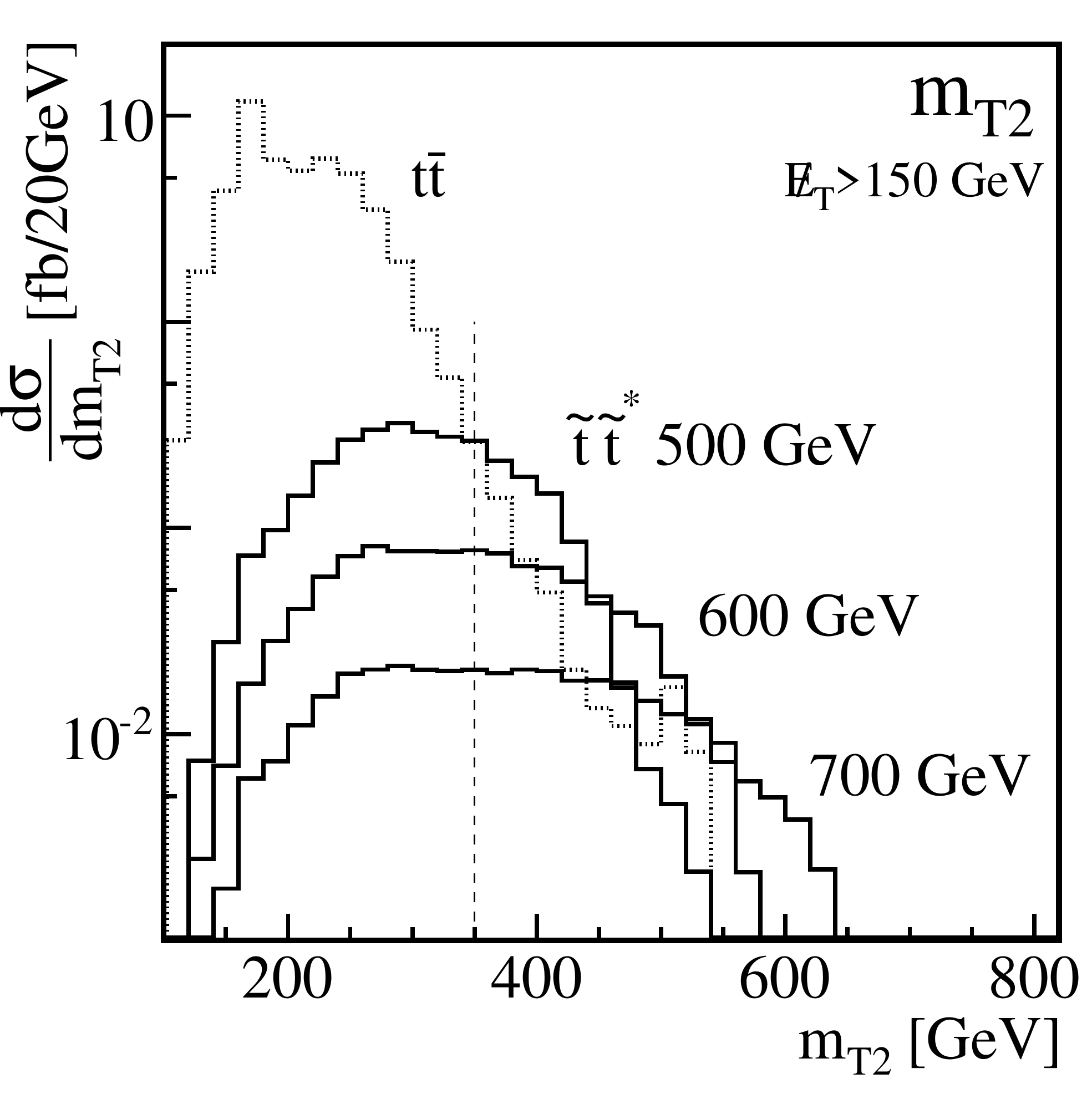}
\caption{$m_{T2}$ distributions for stop and top pair production using 
all (\xpass,\xpass), (\xpass,\xfail), (\xfail,\xpass) and (\xfail,\xfail) events from the
  $b$/jet-bucket algorithm, requiring $p_{T,t}^\text{rec} > 100~\gev$. The distributions for stop pair production with
  stop masses of 400, 500 and 600~GeV are shown after a selection cut on $\slashed{E}_T > 150~\gev$, along with the $t\bar{t}$ background. The vertical dotted line indicates $m_{T2}=350~\gev$. }
\label{fig:mt2}
\end{figure}

\begin{table}[b]
\begin{tabular}{l||r|rrr|rr}
\hline
&$t\bar{t}$+jets [fb]   & \multicolumn{3}{c|}{$\tilde{t} \tilde{t}^\ast$ [fb]} & $S/B$ & $S/\sqrt{B}$ \cr
$m_{\tilde{t}}$ [$\gev$] & & 500 & 600  &  700  & \multicolumn{2}{c}{600}  \cr
\hline
before cuts            & $234\times 10^3$ &    80.50 &     23.00 &      7.19 & & \\
veto lepton & $157\times 10^3$ &   50.45&     14.38&      4.46 & & \\
$\ge 5$ jets &  $85.9 \times 10^3$&     37.87&     10.90&      3.37 & & \\
2 $b$-tags &  $28.0 \times 10^3$&   11.41&      3.30 &      1.02 &   & \\
2 tops reconstructed, $p_{T,t}^\text{rec} > 100$~GeV & $6.90 \times 10^3$  &  4.19 &  1.30 &  0.40 & 0.0002 & 0.08 \cr
$E\!\!\!/_T>150$~GeV & 48.53 &  2.98 &  1.04 &  0.35 & 0.02 & 0.8 \cr
$m_{T2}>350$~GeV & 0.45 &  0.84 &  0.46 &  0.19 & 1.0 & 3.5 \cr
\hline
100\% $\tau$ rejection & 0.12 &  0.77 &  0.42 &  0.17 & 3.6 & 6.1 \cr
\hline
\end{tabular}
\caption{Cross sections for top background and stop pairs with masses of 500, 600, and 700 GeV 
  after selection cuts and application of the $b$/jet bucket
  analysis. We assume exclusively stop decays to 100~GeV
  neutralinos. The significance for 600~GeV stops is given for an integrated luminosity
  of $25~\ifb$.}
\label{tab:stop}
\end{table}

In this section, we assume scalar top pair production followed by
decay into tops and the lightest neutralino $\tilde{\chi}^0_1$ with
100\% branching ratio.  For all model points we set the lightest
neutralino mass to $m_{\tilde{\chi}^0_1}=100~\gev$. Cross sections at
the LHC assuming $\sqrt{s} = 8$~TeV are shown in Table~\ref{tab:stop}.
To generate the signal for stop masses of 500, 600, and 700~GeV we use
{\sc Herwig++}~\cite{herwig}. We normalize the production cross section
to the {\sc Prospino} results at next-to-leading
order~\cite{prospino}.

Since the reconstruction technique described in the previous section
are also applicable for tops from stop decays, we expect good top
reconstruction. To reduce the non-top background we first need to
apply a set of simple selection cuts.  We first require at least five jets, two of them
$b$-tagged.  Then, we require large missing momentum, $\slashed{E}_T >
150$~GeV, and veto isolated leptons.  The results are summarized in
Table~\ref{tab:stop}. Because QCD has no intrinsic
source of missing momentum, and $W$+jets has a small rate and a lepton
we ignore this backgrounds in this paper, and assume mostly $t\bar{t}$ backgrounds
with large missing transverse momentum, typically the result of mismeasurement or $\tau$ decay.\bigskip

Based on the algorithm developed in this paper we require two top
buckets with $b$/jet re-ordering. The two reconstructed bucket momenta
we denote as $p_{t_1}$ and $p_{t_2}$.  After the missing
momentum cut the main background is semi-leptonic top pairs, which
means one of the two tagged tops in the background sample is
mis-tagged.

The advantage of an analysis based on fully hadronic top decays is
that both tops are fully reconstructable~\cite{HEP1,atlas}.  We use
the bucket momenta to compute $m_{T2}(p_{t_1}, p_{t_2},
\slashed{E}_T)$~\cite{mt2}.  Its distributions for the $t\bar{t}$
background and the stop pair signal is shown in Figure~\ref{fig:mt2}.
To extract stop pairs we select events with
\begin{equation}
m_{T2} > 350~\gev \; .
\end{equation}
After this cut and for a stop mass of 600~GeV we arrive at $S/B \sim
1$ and more than three sigma significance at the 8~TeV LHC with the
currently available integrated luminosity of $25~\ifb$. In addition, the endpoint
of the $m_{T2}$ distribution with fully reconstructed hadronic tops
should allow us to precisely measure the stop mass~\cite{HEP1}. All
intermediate steps as well as results for other stop masses are shown
in Table~\ref{tab:stop}. Note that some numbers are different from those 
shown in Table~\ref{tab:number_pt100} due to the leptonic decays. \bigskip

Of all events with two reconstructed tops about 10\% involve $\tau$
leptons, both for the signal and the background.  After the missing
momentum cut a significant fraction ($\sim 75\%$) of the top background comes from these
events. In contrast, only 10\% of the signal events include a top decay
to a $\tau$.  Therefore, a $\tau$-rejection would improve our results
significantly, as shown in Table~\ref{tab:stop}.


\section{Conclusion}
\label{sec:conclusion}

In this paper we have presented a new method to identify and
reconstruct hadronically decaying top quarks. It is based on assigning
regular jets to buckets, one for each top decay and one for initial
state radiation. The buckets corresponding to tops are each seeded
with one of the two $b$-jets we require in every events.  If a top
bucket includes all three top decay products it has to fulfill $W$ and
top mass constraints. However, frequently the softer $W$ decay jet is
missing, so we have to rely on the two leading jets to reconstruct 
a defined fraction of the top mass. After an appropriate re-ordering
of the buckets missing the softest decay jets both kinds of buckets
can be used to reconstruct the top four-momentum.\bigskip

To suppress tops which for one or another reason cannot be matched to
a generated top quark we apply a self consistency condition (QMM) to
each bucket. This condition defines the lower bound of the typical
transverse momentum range $100~\gev < p_{T,t} < 350~\gev$ to which the
method is sensitive.  For higher boosts the buckets will eventually
fail due to the size of the jets they are constructed from. For top
quarks with this moderate boost we achieve a maximum efficiency around
60-70\% for the reconstruction of two top quarks. In particular, for
$p_{T,t} < 250$~GeV our method gives a significant improvement over
subjet-based top taggers, which have low efficiencies in this
regime.\bigskip

To illustrate our approach in a new physics framework we have applied
it to supersymmetric stop searches, relying on stop decays to tops and
missing energy. Because we reconstruct the top four-momenta we can
apply a simple $m_{T2}$ analysis, including a measurement of the stop
mass. This makes stop search strategies as simple as sbottom or
slepton searches.\bigskip

While the detailed numerical results for our method should be tested
in a realistic experimental environment there obviously exists a wide
range of possible applications for top buckets in ATLAS and CMS. As a
first step, hadronic top pair production with and without
contributions from beyond the Standard Model might serve as a useful
testing ground~\cite{atlas}.

\acknowledgments

We would like to thank the Aspen Center of Physics because the idea
for this paper was born on a Snowmass ski lift.
Fermilab is operated by Fermi Research Alliance, LLC, under contract DE-AC02-07CH11359 with the United States Department of Energy. MRB would like to thank Joseph Lykken and Maria Spiropulu for useful advice. MT would like to thank Bobby Acharya for helpful discussions.


\begin{thebibliography}{99}

\bibitem{tops_ex}
 see \eg
 F.~-P.~Schilling,
  Int.\ J.\ Mod.\ Phys.\ A {\bf 27}, 1230016 (2012).

  
\bibitem{higgs}
G.~Aad {\it et al.} [ATLAS Collaboration],
 Phys.\ Lett.\ B {\bf 716}, 1 (2012);
S.~Chatrchyan {\it et al.}  [CMS Collaboration],
 Phys.\ Lett.\ B {\bf 716}, 30 (2012),

\bibitem{lecture}
 for a pedagogical introduction see 
  T.~Plehn,
  Lect.\ Notes Phys.\  {\bf 844}, 1 (2012)
  [arXiv:0910.4182].

\bibitem{bsm_review}
  D.~E.~Morrissey, T.~Plehn and T.~M.~P.~Tait,
  Phys.\ Rept.\  {\bf 515}, 1 (2012).

\bibitem{seymour}
 M.~H.~Seymour,
  Z.\ Phys.\  C {\bf 62}, 127 (1994);

\bibitem{tagger_review}
 T.~Plehn and M.~Spannowsky,
  J.\ Phys.\ G {\bf 39}, 083001 (2012);
  A.~Abdesselam {\it et al.},
  Eur.\ Phys.\ J.\ C {\bf 71}, 1661 (2011);
  A.~Altheimer {\it et al.},
  J.\ Phys.\ G {\bf 39}, 063001 (2012).

\bibitem{early_toptagger}
 B.~Holdom,
  JHEP {\bf 0703}, 063 (2007);
 W.~Skiba and D.~Tucker-Smith,
  Phys.\ Rev.\  D {\bf 75}, 115010 (2007);
 M.~Gerbush, T.~J.~Khoo, D.~J.~Phalen, A.~Pierce and D.~Tucker-Smith,
   Phys.\ Rev.\  D {\bf 77}, 095003 (2008);
 J.~Thaler and L.~-T.~Wang,
  JHEP {\bf 0807}, 092 (2008);
 D.~E.~Kaplan, K.~Rehermann, M.~D.~Schwartz and B.~Tweedie,
  Phys.\ Rev.\ Lett.\  {\bf 101}, 142001 (2008);
 L.~G.~Almeida, S.~J.~Lee, G.~Perez, G.~F.~Sterman, I.~Sung and J.~Virzi,
  Phys.\ Rev.\ D {\bf 79}, 074017 (2009).

\bibitem{bdrs}
  J.~M.~Butterworth, A.~R.~Davison, M.~Rubin and G.~P.~Salam,
  Phys.\ Rev.\ Lett.\  {\bf 100}, 242001 (2008).

\bibitem{tth}
 T.~Plehn, G.~P.~Salam and M.~Spannowsky,
  Phys.\ Rev.\ Lett.\  {\bf 104}, 111801 (2010).

\bibitem{HEP1}
 T.~Plehn, M.~Spannowsky, M.~Takeuchi, and D.~Zerwas,
  JHEP {\bf 1010}, 078 (2010);
 \url{http://www.thphys.uni-heidelberg.de/~plehn}

\bibitem{atlas}
 G.~Aad {\it et al.} [ATLAS Collaboration],
  ATLAS-CONF-2012-102;
 G.~Aad {\it et al.}  [ATLAS Collaboration],
  JHEP {\bf 1301}, 116 (2013);
 Gregor Kasieczka, PhD thesis, Heidelberg University, to appear.

\bibitem{HEP3}
  T.~Plehn, M.~Spannowsky and M.~Takeuchi,
  Phys.\ Rev.\ D {\bf 85}, 034029 (2012).

\bibitem{alpgen}
 M.~L.~Mangano, M.~Moretti, F.~Piccinini, R.~Pittau and A.~D.~Polosa,
  JHEP {\bf 0307}, 001 (2003).
  
\bibitem{pythia}
 T.~Sjostrand, S.~Mrenna and P.~Z.~Skands,
  JHEP {\bf 0605}, 026 (2006);
 T.~Sjostrand, S.~Mrenna and P.~Z.~Skands,
  Comput.\ Phys.\ Commun.\  {\bf 178}, 852 (2008).

\bibitem{mlm}
 M.~L.~Mangano, M.~Moretti, F.~Piccinini and M.~Treccani,
  JHEP {\bf 0701} (2007) 013.

\bibitem{top_nnlo}
  N.~Kidonakis,
  Phys.\ Rev.\ D {\bf 82}, 114030 (2010);
 V.~Ahrens, A.~Ferroglia, M.~Neubert, B.~D.~Pecjak and L.~L.~Yang,
  Phys.\ Lett.\ B {\bf 703}, 135 (2011);
 M.~Cacciari, M.~Czakon, M.~L.~Mangano, A.~Mitov and P.~Nason,
  Phys.\ Lett.\ B {\bf 710}, 612 (2012);
 S.~Moch, P.~Uwer and A.~Vogt,
  arXiv:1203.6282 [hep-ph].

\bibitem{ca_algo}
 Y.~L.~Dokshitzer, G.~D.~Leder, S.~Moretti and B.~R.~Webber,
  JHEP {\bf 9708}, 001 (1997);
 M.~Wobisch and T.~Wengler,
  arXiv:hep-ph/9907280.

\bibitem{fastjet}
 M.~Cacciari and G.~P.~Salam,
  Phys.\ Lett.\  B {\bf 641}, 57 (2006);
 M.~Cacciari, G.~P.~Salam and G.~Soyez,
  Eur.\ Phys.\ J.\ C {\bf 72}, 1896 (2012);
  \url{http://fastjet.fr}
 
\bibitem{endpoint}
 T.~Plehn and M.~Takeuchi,
  J.\ Phys.\ G {\bf 38}, 095006 (2011).

\bibitem{stops_ex}
 for results based on 8~TeV data see \eg
  ATLAS-CONF-2012-166, 
  ATLAS-CONF-2012-167, 
  ATLAS-CONF-2013-001, 
  CMS-PAS-SUS-12-023.


\bibitem{stops_semilep}
 P.~Meade and M.~Reece,
  Phys.\ Rev.\ D {\bf 74}, 015010 (2006);
 T.~Han, R.~Mahbubani, D.~G.~E.~Walker and L.~-T.~Wang,
  JHEP {\bf 0905}, 117 (2009);
 Y.~Bai, H.~-C.~Cheng, J.~Gallicchio and J.~Gu,
  JHEP {\bf 1207}, 110 (2012);
 Z.~Han, A.~Katz, D.~Krohn and M.~Reece,
  JHEP {\bf 1208}, 083 (2012);
 C.~Kilic and B.~Tweedie,
  arXiv:1211.6106 [hep-ph].

\bibitem{stops_had}
 D.~S.~M.~Alves, M.~R.~Buckley, P.~J.~Fox, J.~D.~Lykken and C.~-T.~Yu,
  arXiv:1205.5805 [hep-ph];
 D.~E.~Kaplan, K.~Rehermann and D.~Stolarski,
  JHEP {\bf 1207}, 119 (2012);
 B.~Dutta, T.~Kamon, N.~Kolev, K.~Sinha and K.~Wang,
  Phys.\ Rev.\ D {\bf 86}, 075004 (2012).

\bibitem{stops_hep}
 T.~Plehn, M.~Spannowsky and M.~Takeuchi,
  JHEP {\bf 1105}, 135 (2011)
  and
  JHEP {\bf 1208}, 091 (2012).

\bibitem{herwig}
 G.~Corcella {\it et al.},
  arXiv:hep-ph/0210213;
 M.~Bahr \etal,
  arXiv:0812.0529 [hep-ph].

\bibitem{prospino}
 W.~Beenakker, M.~Kr\"amer, T.~Plehn, M.~Spira and P.~M.~Zerwas,
  Nucl.\ Phys.\  B {\bf 515}, 3 (1998); \\
 for further improvements see \eg
 W.~Hollik, M.~Kollar and M.~K.~Trenkel,
  JHEP {\bf 0802}, 018 (2008);
 A.~Kulesza and L.~Motyka,
  Phys.\ Rev.\ D {\bf 80}, 095004 (2009);
 W.~Beenakker, S.~Brensing, M.~Kramer, A.~Kulesza, E.~Laenen and I.~Niessen,
  JHEP {\bf 1008}, 098 (2010);
 M.~Beneke, P.~Falgari and C.~Schwinn,
  Nucl.\ Phys.\ B {\bf 842}, 414 (2011).

\bibitem{mt2}
 C.~G.~Lester and D.~J.~Summers,
  Phys.\ Lett.\  B {\bf 463}, 99 (1999);
 A.~Barr, C.~Lester and P.~Stephens,
  J.\ Phys.\ G {\bf 29}, 2343 (2003).

\end{thebibliography}
\end{document}

\bibitem{HEP_ATLAS}
ATLAS-CONF-2012-065,
\url{https://atlas.web.cern.ch/Atlas/GROUPS/PHYSICS/CONFNOTES/ATLAS-CONF-2012-065}